\documentclass[journal]{IEEEtran}

\usepackage{balance}
\usepackage{graphicx}        
\usepackage{amssymb}
\usepackage{algorithm}
\usepackage{algpseudocode}
\usepackage{amsmath,todonotes}
\usepackage{mathtools}
\usepackage{comment}
\usepackage{multirow}
\usepackage{booktabs}
\usepackage{cite}

\newtheorem{problem}{Problem}
\newtheorem{proposition}{Proposition}
\newtheorem{lemma}{Lemma}
\newtheorem{remark}{Remark}

\newtheorem{assumption}{Assumption}
\newtheorem{theorem}{Theorem}
\newtheorem{corollary}{Corollary}
\newtheorem{definition}{Definition}
\newtheorem{example}{Example}

\title{\LARGE \bf Actuator Security Indices Based on Perfect Undetectability: Computation, Robustness, and Sensor Placement}

\author{Jezdimir Milo\v{s}evi\'{c}$^{1}$, Andr\'{e} Teixeira$^{2}$, Karl H. Johansson$^{1}$, and Henrik Sandberg$^{1}$
\thanks{$^{1}$Division of Decision and Control Systems, School of
Electrical Engineering and Computer Science, KTH Royal Institute of Technology, Stockholm, Sweden. Emails: \{jezdimir, hsan, kallej\}@kth.se.
$^{2}$Signals and Systems, Department of Engineering Sciences, Uppsala University, Uppsala, Sweden. Email: \{andre.teixeira\}@angstrom.uu.se.       
        }
}

\begin{document}

\maketitle

\begin{abstract}
This paper proposes an actuator security index based on the definition of perfect undetectability. 
This index can help a control system operator to localize the most vulnerable actuators in a networked control system, which can then be secured. 
 Particularly, the security index of an actuator equals the minimum number of sensors and actuators that needs to be compromised, such that a perfectly undetectable attack against that actuator can be conducted. 
 A method for computing the index for small scale networked control systems is derived, and it is shown that the index can potentially be increased by placing additional sensors. 
 The difficulties that appear once the system is of a large scale are then outlined: the problem of calculating the index is NP--hard, the index is vulnerable to system variations, and it is based on the assumption that the attacker knows the entire model of the system. 
To overcome these difficulties, a robust security index is introduced. 
The robust index can be calculated in polynomial time, it is unaffected by the system variations, and it can be related to both limited and full model knowledge attackers. 
Additionally, we analyze two sensor placement problems with the objective to increase the robust indices.
We show that both of these problems have submodular structures, so their suboptimal solutions with performance guarantees can be obtained in polynomial time.  
Finally,  the theoretical developments are illustrated through numerical examples.   
\end{abstract}

\section{Introduction}

Actuators are one of the most vital components of networked control systems.
Through them, we ensure that important physical processes such as power production or water distribution behave in a desired way. 
Actuators can also be expensive, so it is important to carefully choose where to place them.
To solve this important problem of cost--efficient allocation of actuators, number of approaches have been developed~\cite{7039832,6762966,7122316,clark2012leader}. 
However, an issue with these approaches is that they do not take security aspects into consideration. 
This could be dangerous, since control systems can easily become a target of malicious adversaries~\cite{slay2007lessons,stuxnet,case2016analysis}. 
For this reason, it is essential to be able to check if these efficient actuator placements are at the same time secure. 

Motivated by this issue, we introduce novel actuator security indices $\delta$ and $\delta_r$.
As we shall see, these indices can be used for both localization of vulnerable actuators, and for  development of defense strategies.  
The security index $\delta(u_i)$ is defined for every actuator $u_i$, and it is equal to the minimum number of sensors and actuators that needs to be compromised by an attacker to conduct a perfectly undetectable attack against $u_i$. 
Perfectly undetectable attacks are dangerous, since they do not leave any trace in the sensor measurements~\cite{6816560,7479526}.
Therefore, an actuator with a small value of $\delta$ is potentially very vulnerable. 
Since $\delta$ cannot be straightforwardly used in large scale networked systems, as explained in this paper,
we introduce the robust security index $\delta_r$ as its replacement in these systems.
We then outline favorable properties of $\delta_r$, and propose possible strategies for increasing $\delta_r$. 
Finally, we remark that due to the properties of perfectly undetectable attacks, sensor security indices can in general be derived from actuator security indices. 
Hence, the focus of the paper is exclusively on actuator indices. 

\textit{Literature Review. }It has been recognized within the control community that cyber-attacks  require new techniques to be handled~\cite{cardenas2008secure}. 
For instance, cyber-attacks impose fundamental limitations for state estimation~\cite{tabuada,7322210}, detection~\cite{6545301}, and for consensus computation~\cite{4586680,5779706}. 
The most troublesome attacks are those that can inflict considerable damage, while remaining unnoticed by the system operator. Examples include stealthy false-data injection attacks~\cite{liustatic}, undetectable (zero-dynamics) attacks~\cite{6545301,bias}, perfectly undetectable attacks~\cite{6816560,7479526}, covert attacks~\cite{7011176}, optimal linear attacks~\cite{kalelinearattacks}, and replay attacks~\cite{mo2015physical}. 
 To characterize the vulnerability of the system and protect it against these attacks, many different approaches have been proposed~\cite{lun2016cyber,7935369,7011179}.
 
In this work, we focus on so--called security indices. 
The first security index was introduced in~\cite{sandberg2010security}. 
In this work, a static linear system was used as a network model, and the static security index $\alpha$ was defined for each sensor. 
The main purpose of this index is to help the operator to localize the most vulnerable sensors in a power network, which are those with low values of $\alpha$.
Once these sensors are localized, the operator can allocate additional security measures to protect them.
Furthermore, to choose the most beneficial combination of security measures, he/she can again use security indices~\cite{secalo3}.

A major challenge is to compute $\alpha$ once the power network is large.
In fact, it was shown that the problem of calculating~$\alpha$ is NP--hard in general~\cite{hendrickx2014efficient}. 
However, $\alpha$ can be efficiently computed in some cases~\cite{6160456,sou2012computing,hendrickx2014efficient,kosut2014max,yamaguchi2015cyber}. 
For instance, ~\cite{6160456} proposes an upper bound on $\alpha$. 
This bound can be obtained in polynomial time by solving the minimum $s$--$t$ cut problem, and proves to be tight in several cases of interest.

Although $\alpha$ proved to be a useful tool for both vulnerability analysis and development of defense strategies, 
security indices that can be used for more general dynamical systems have been considered only by few works~\cite{7799178,7580001}. 
The index in~\cite{7799178} considerably differs from $\alpha$, since it characterizes vulnerability of the entire system. 
On the other hand, in~\cite{7580001}, the definition of \textit{undetectability}~\cite{6545301} was used to define a security index similar to $\alpha$ to characterize vulnerability of sensors and actuators within the system.
However, this work neither addresses the problems that appear in large scale systems, nor explains how this index can be used for defense purposes.
In this paper, we introduce novel actuator security indices suitable for dynamical systems, tackle the challenges that appear once the system is of a large scale, and propose defense strategies based on these indices.

\textit{Contributions. }The contributions of this manuscript are as follows.  
Firstly, we propose a novel type of actuator security index $\delta$.
In contrast to the dynamical index proposed in~\cite{7580001}, our index is based on the definition of \textit{perfect undetectability}~\cite{6816560,7479526}. 
To calculate $\delta$ when the number of sensors and actuators is small, we derive a sufficient and necessary condition that a set of compromised components needs to satisfy in order for a solution of the security index problem to exist (Proposition~\ref{theorem:condition}).  
To prove Proposition~\ref{theorem:condition}, we use an algebraic condition for existence of perfectly undetectable attacks derived in~\cite{6816560,7479526}.  
We also show that $\delta$ can potentially be increased by placing additional sensors, and that placement of additional actuators may decrease $\delta$ (Proposition~\ref{prop:adding_sensors_delta}).
We then identify the three issues that appear once the system is of a large scale: 
(1)~The problem of computing $\delta$ is NP--hard (Theorem~\ref{theorem:NPhardness}); 
(2)~$\delta$ is fragile to system variations, which are expected in large systems;
(3)~$\delta$ is based on the assumption that the attacker knows the entire model of the system, which can be a conservative assumption in this case. 

To overcome these deficiencies, we introduce the robust security index $\delta_r$, which is our second contribution. 
To define~$\delta_r$, we use a structural model of the system~\cite{dion2003generic}, and the notion of vertex separators that was used to characterize existence of perfectly undetectable attacks in~\cite{7479526}. 
Particularly, we first show how vertex separators can be used to upper bound the index~$\delta$ (Theorem~\ref{proposition:sec_index_sufficient_condition}), and then define $\delta_r$ to be the best upper bound based on vertex separators. 

Thirdly, we show that $\delta_r$ does not suffer from the aforementioned deficiencies of $\delta$. 
Namely, $\delta_r$ can be calculated efficiently by solving the minimum $s$--$t$ cut problem in a graph (Proposition~\ref{theorem:security_index_ts_cut}). 
We remark that Proposition~\ref{theorem:security_index_ts_cut} extends the previous work on the static index $\alpha$~\cite{6160456,sou2012computing,hendrickx2014efficient,kosut2014max}, where the minimum $s$--$t$ cut problem was also used for calculating/approximating $\alpha$.  
Additionally, $\delta_r$ is unaffected by the system variations, since it is based on the structural model of the system~\cite{dion2003generic}. 
Moreover, $\delta_r$ can be related to both full and limited model knowledge attackers.
In the context of the full model knowledge attacker, $\delta_r(u_i)$ characterizes the minimum amount of resources for conducting a perfectly undetectable attack against $u_i$ in any possible realization of the system (Proposition~\ref{prop:FMNA_resources}). 
We then introduce an attacker with resources limited to a local model and measurements, 
and prove that he/she can also conduct a perfectly undetectable attack against $u_i$ by compromising a right combination of $\delta_r(u_i)$ components (Proposition~\ref{theorem:restricted_attacker}). 
We also analyze an attacker that knows only the structural model of the system.
In this case, $\delta_r(u_i)$ lower bounds the number of components this attacker needs to compromise to ensure that the attack against $u_i$ remains perfectly undetectable (Proposition~\ref{prop:WMNA_resources}).

Since the previous results imply that actuators with small value of $\delta_r$ are potentially very vulnerable, we propose sensor placement strategies to increase $\delta_r$, which we outline as our fourth contribution. 
We firstly show that $\delta_r$ is guaranteed to increase if sensors are placed to suitable locations in the system (Thereom~\ref{proposition:place_to_put_sensors}). 
Based on this result, we formulate two sensor placement problems with the objective to increase $\delta_r$, and show that these problems have suitable submodular structures (Proposition~\ref{theorem:improving_bound_subm_2}--\ref{theorem:improving_bound_subm_1}).
This enables us to find suboptimal solutions of these problems with guaranteed performance efficiently, even in large scale networked control systems. 
Finally, we illustrate the theoretical results through numerical examples. 

The preliminary version of the paper appeared in~\cite{jezdimir_allerton}. 
This work differs from~\cite{jezdimir_allerton} in the following aspects: 
(1)~We prove that $\delta$ is NP--hard to calculate (Theorem~\ref{theorem:NPhardness}); 
(2)~The connection of $\delta_r$ with the full/limited model knowledge attacker is derived (Propositions~\ref{prop:FMNA_resources}--\ref{prop:WMNA_resources}); 
(3)~We prove that both $\delta$ and $\delta_r$ can be increased by placing additional sensors (Proposition~\ref{prop:adding_sensors_delta}, Theorem~\ref{proposition:place_to_put_sensors}); 
(4)~A new section on increasing $\delta_r$ is added (Section~\ref{section:improving_upper_bound}); 
(5)~More detailed proofs of the results that appeared in~\cite{jezdimir_allerton} are included (Proofs of Propositions~\ref{theorem:condition} and~\ref{theorem:security_index_ts_cut}, and Theorem~\ref{proposition:sec_index_sufficient_condition}); 
(6)~The section with examples is extended.

\textit{Organization. }
The remainder of the paper is organized as follows. 
In Section~\ref{section:Model_Setup}, we introduce the system model, the attacker model, and the security index $\delta$. 
In Section~\ref{section:properties}, we investigate properties of~$\delta$. 
In Section~\ref{section:upper_bound}, we derive an upper bound on $\delta$, and based on it, define the robust index $\delta_r$. 
In Section~\ref{section:upper_bound_properties}, we outline properties of $\delta_r$. 
In Section~\ref{section:improving_upper_bound}, we discuss strategies for increasing $\delta_r$. 
In Section~\ref{section:simulations}, we illustrate the theoretical findings through examples. 
In Section~\ref{section:conclusion}, we conclude. 
Appendix contains the proofs of the results. 

\section{Security Index $\delta$\label{section:Model_Setup}} 

In this section, we introduce the model setup and formulate the problem of calculating the actuator security index $\delta$. 
We remark that although we consider discrete time systems, the analysis presented in the paper can also be extended to continuous time systems.

\subsection{Model Setup}

The plant of a 
 networked control system is modeled by
\begin{equation}
\begin{aligned}
x(k+1)&=Ax(k)+B u(k)+B_a a(k),\\
y(k)&=C x(k)+D_a a(k),
\label{eqn:systemg}
\end{aligned}
\end{equation}
where $x(k)$$ \in$$ \mathbb{R}^{n_x}$ is the system state at time step $k$$ \in$$ \mathbb{Z}_{\geq 0}$, $u(k)$$\in $$\mathbb{R}^{n_u}$ is the control input, $y(k) $$\in $$\mathbb{R}^{n_y+n_e}$ is the measurement vector, and $a(k)$$\in $$\mathbb{R}^{n_u + n_y}$ is the attack vector.
For the analysis that follows, it is convenient to assume that the system is in a steady state $x(0)$$=$$0$ and $u$$=$$0$\footnote{For a signal $s: \mathbb{Z}_{\geq 0} \rightarrow \mathbb{R}^{n_s}$, $s$$ =$$ 0$ means that $s(k)$$ =$$ 0$ for all $k $$\in$$ \mathbb{Z}_{\geq 0}$, while $s $$\neq $$0$ means $s(k)$$ \neq $$0$ for at least one $k $$\in$$ \mathbb{Z}_{\geq 0}$. }. 
Due to linearity, this assumptions is without loss of generality for most results in the paper.
The exceptions are clearly outlined. 
We also allow the last $n_e $$\geq$$ 0$ elements of $y$ to be protected, so the attacker cannot manipulate them. 
The protection can be achieved by implementing encryption/authentication schemes, and/or improving physical protection~\cite{secalo3}.

We now introduce the attacker model.
The first $n_u$ elements of $a$ model attacks against the actuators, while the last $n_y$ model attacks against the unprotected sensors. 
The matrices $B_a$ and $D_a$ are therefore given by
\begin{equation*}
B_a=\begin{bmatrix} B & \textbf{0}_{n_x \times n_y} \end{bmatrix}, \hspace{5mm} D_a=\begin{bmatrix} \textbf{0}_{n_y \times n_u} & \textbf{I}_{n_y} \\ 
\textbf{0}_{n_e \times n_u} & \textbf{0}_{n_e \times n_y}
\end{bmatrix},
\end{equation*} 
and $B$ is assumed to have a full column rank.
This is needed to exclude degenerate cases in which the attacks trivially cancel each-other, or cases where an actuator does not affect the system. 
We denote by $\mathcal{I}$$=$$\{1,\ldots,n_u+n_y\}$ the indices of elements of $a$, and by $a^{(\mathcal{I}_a)}(k)$ the vector consisting of the elements of $a(k)$ with indices from $\mathcal{I}_a$$ \subseteq $$\mathcal{I}$.
The set $\mathcal{I}$ is also used to denote the joint set of actuators and unprotected sensors in the first part of the paper. 
We adopt the following common assumption about the attacker. 

\begin{assumption}\label{assumption:attacker_full_model_knowledge}
The attacker: (1) Can read and change the values of attacked control signals and measurements $\mathcal{I}_a $$\subseteq $$\mathcal{I}$ arbitrarily; (2) Knows $A,B,C$.  
\end{assumption}

Further, we assume that the attacker's goal is to conduct an attack while ensuring the attack remains undetected by the system operator.
To model this goal, we need a suitable definition of undetectability. 
In this paper, we use the definition of \textit{perfect undetectability}~\cite{6816560,7479526}.

\begin{definition} \label{definition:perfectly_undetectable}
Let $x(0)$$=$$0$ and $u$$=$$0$. The attack signal $a$$ \neq$$ 0$ is \textit{perfectly undetectable} if $y$$=$$0$. 
\end{definition}

In other words, the attack is perfectly undetectable if it does not leave any trace in the sensor measurements. 
For this reason, these attacks are potentially very dangerous.

\subsection{Security Index $\delta$: Problem Formulation}

We now introduce an actuator security index $\delta$.  
The security index $\delta(i)$ is defined for every actuator $i \in \mathcal{I}$.
The index is equal to the minimum number of sensors and actuators that need to be compromised by the attacker, such as to conduct a perfectly undetectable attack.
Additionally, $i$ has to be actively used in the attack, which models a goal or intent by the attacker. 
Naturally, actuators with small values of $\delta$ are more vulnerable than those with large values. 
In the worst case, $\delta(i)$$=$$1$. 
This implies that an attacker can attack~$i$ and stay perfectly undetectable without compromising any other component. 
Let  $||a||_{0}=|\cup_{k \in \mathbb{Z}_{\geq 0}} {\rm supp}(a(k))|,$ 
where $\text{supp}(a(k))=\{i\in \mathcal{I}:a^{(i)}(k) \neq 0\}$. Based on the previous discussion, $\delta(i)$ can be formally defined as follows.

\begin{problem}\label{problem:sec_index_perfectly_undetectable} \textit{Calculating $\delta$}
\begin{align*}
\underset{a}{\text{minimize}}\hspace{12mm} \delta(i)&=||a||_{0}\\
\text{subject to} \hspace{4mm} x(k+1)&=Ax(k)+B_a a(k),  &&\text{(C1)}\\
    0&=Cx(k)+D_a a(k),   &&\text{(C2)} \\
    x(0)&=0,   &&\text{(C3)}\\
    a^{(i)} &\neq 0. && \text{(C4)}
\end{align*}
\end{problem}
The objective function reflects our desire to find the minimum number of sensors and actuators to conduct a perfectly undetectable  attack (sparsest signal $a: \mathbb{Z}_{\geq 0} \rightarrow \mathbb{R}^{n_u+n_y}$). 
The constraints: (C1) and (C2) ensure that the attack signal satisfies physical dynamics of the system; (C2) and (C3) constraint the attack to be perfectly undetectable; (C4) ensures that actuator $i$ is actively used in the attack.

Before we start analyzing $\delta$,
we point out several properties of Problem~\ref{problem:sec_index_perfectly_undetectable}.
Firstly, this problem is not necessarily feasible for every actuator $i$. 
Absence of a solution implies that the attacker cannot attack $i$ while staying perfectly undetectable.  
Thus, we adopt $\delta(i)$$=$$+\infty$ in this case. 
Secondly,  if we remove (C3) and include $x(0)$ to be an optimization variable, we recover the security index problem based on undetectable attacks~\cite{7580001}. 
Thirdly, Problem~\ref{problem:sec_index_perfectly_undetectable} can also be used for finding security indices of unprotected sensors. 
However, to conduct a perfectly undetectable attack, at least one actuator must be attacked to make the attack signal against a sensor active. 
Thus, the problem of finding $\delta(i)$ of sensor $i \in \mathcal{I}$ can in general be reduced to the problem of finding an actuator with the minimum $\delta$ that excites sensor~$i$.  
Finally, the problem can also be extended to capture the case where sensors and actuators are not equally hard to attack. 

 \section{Properties of $\delta$~\label{section:properties}}

We now analyze properties of $\delta$.
We show how $\delta$ can be computed once $\mathcal{I}$ has small cardinality, and that $\delta$ can be increased by placing additional sensors.
We then outline difficulties that appear in large scale networked control systems:  Problem~\ref{problem:sec_index_perfectly_undetectable} is NP--hard, $\delta$ can be quite vulnerable to system variations, and  Assumption~\ref{assumption:attacker_full_model_knowledge}.(2) may be conservative in this case. 
Overall, $\delta$ is more appropriate for small scale systems, while a replacement is required for large scale systems. 
Proofs of the results from this section are available in Appendix~\ref{appendix:proofs_1}. 

\subsection{Calculating $\delta$ Using Brute Force Search} 

 We first derive a sufficient and necessary condition that the set of attacked components $\mathcal{I}_a$ needs to satisfy, so that we can construct an attack signal $a$ feasible for Problem~\ref{problem:sec_index_perfectly_undetectable}. 
We then explain how this condition can be used for finding $\delta$.
Prior to that, we introduce some terminology and notation.  
The transfer function from $a$ to $y$ is denoted by $G$, and the \textit{normal rank} of $G$ is defined as $\text{normrank }G =\max\{\text{rank } G(z) |z\in \mathbb{C}\}.$   
With $G^{(\mathcal{I}_a)}$, we denote the transfer function matrix that contains the columns of $G$ from $\mathcal{I}_a \subseteq \mathcal{I}$.

\begin{proposition} \label{theorem:condition}
A perfectly undetectable attack conducted with components $\mathcal{I}_a $$\subseteq$$ \mathcal{I}$ in which component $i$$\in$$ \mathcal{I}_a$ is actively used exists if and only if
\begin{equation} \label{eqn: condition_normal_ranks} 
\text{normrank }  G^{(\mathcal{I}_a)} = \text{normrank } G^{(\mathcal{I}_a \setminus i)}. 
\end{equation}
\end{proposition}

There are two important consequences of this result. 
Firstly, we can use~\eqref{eqn: condition_normal_ranks} to calculate  $\delta(i)$ of actuator $i$ in small scale systems in the following way.  
We form all the combinations of sensors and actuators $\mathcal{I}_a \subseteq \mathcal{I}$, $i \in \mathcal{I}_a$, of cardinality $p$.
The initial value of $p$ is set to $1$. 
For each combination, we check if~\eqref{eqn: condition_normal_ranks} is satisfied, which can be done efficiently (e.g. by using the Matlab function \texttt{tzero}).
If we find a combination that satisfies~\eqref{eqn: condition_normal_ranks}, we stop the search.
The value of $\delta(i)$ is then $p$.
If~\eqref{eqn: condition_normal_ranks} is not satisfied for any of the combinations of cardinality $p$, we increase $p$ by 1, and repeat the process.

Secondly, as shown in the proof,  the attacker can perfectly cover an arbitrarily large attack signal injected in $i$ once~\eqref{eqn: condition_normal_ranks} holds. 
Additionally, he/she can construct this attack off-line using only the model knowledge, 
which makes the attack decoupled from $x(0)$ and $u$.
Thus, the attack remains perfectly undetectable for any choice of $x(0)$ and $u$, and the assumption $x(0)=0$ and $u=0$ is without lose of generality in this case. 
However, the attack is implemented in a feedforward manner, which makes it fragile in respect of  modeling errors~\cite{teixeira2012revealing}. 
We further discuss these properties in Section~\ref{section:simulations}.

\subsection{Increasing $\delta$}

We now investigate how the deployment of new sensors and actuators affects $\delta$.

\begin{proposition} \label{prop:adding_sensors_delta}
Assume that a new component $j$ (sensor or actuator) is deployed. Let $\delta(i)$ and $\delta'(i)$ be respectively the security indices of an arbitrary actuator $i$ before and after the deployment. Then: 
(1) $\delta(i) $$\leq$$ \delta'(i)$$ \leq $$\delta(i)+1$ if $j$ is an unprotected sensor; (2) $\delta(i) \leq \delta'(i) $ if $j$ is a protected sensor; (3) $\delta(i) \geq \delta'(i) $ if $j$ is an actuator.
\end{proposition}

Proposition~\ref{prop:adding_sensors_delta} has two interesting consequences. 
Firstly, it implies that we can increase $\delta$ by placing additional sensors to monitor the system. 
Furthermore, $\delta$ can be used to determine which sensor placement is the most beneficial. 
For example, one optimality criterion can be to select the placement such that the minimum value of $\delta$ is as large as possible. 
If the system is of a small scale, and if a small number of sensors is being placed, we can simply go through all the combinations of sensors and pick the best.  
Secondly, Proposition~\ref{prop:adding_sensors_delta} illustrates an interesting trade-off between security and safety. 
On the one hand, to make the system easier to control and more resilient to actuator faults, more actuators should be placed in the system.
On the other, this may also decrease the security indices, so the actuators become easier to attack.

\subsection{$\delta$ and Large Scale Networked Control Systems}
We now outline difficulties that appear once a networked control system is of a large scale. 
\subsubsection{NP Hardness of Problem~\ref{problem:sec_index_perfectly_undetectable}} 
We showed earlier that $\delta$ can in general be obtained by using the brute force search. 
However, this method is computationally intense, and it is inapplicable for large scale networked systems. 
In fact, Theorem~\ref{theorem:NPhardness} that we introduce next establishes that Problem~\ref{problem:sec_index_perfectly_undetectable} is NP-hard.
Thus, there are no known polynomial time algorithms that can be used to solve this problem. 

\begin{theorem} \label{theorem:NPhardness}
Problem~\ref{problem:sec_index_perfectly_undetectable} is NP-hard. 
\end{theorem}

\subsubsection{Fragility of $\delta$}
Large scale networked control systems are complex systems that can change configuration over time.
For example, in the power grids, micro-grids can detach from the grid~\cite{SIMPSONPORCO20132603}, some of the power lines may be turned--off~\cite{amin2007preventing}, or some measurements may become unavailable due to unreliable communication~\cite{IMER20061429}.   
Unfortunately, the security index can be quite fragile with respect to changes in realization of the system matrices $A,B,C$, as shown in the following example.

\begin{example} \label{ex:example_1}  Let the realization of the system be
\begin{equation*}
\begin{aligned}
A=\begin{bmatrix} 0.1 & 0 \\ 0.01 & 0.1  \end{bmatrix}, \hspace{1mm} B=\begin{bmatrix} 1 \\ 0 \end{bmatrix}, \hspace{1mm}C=\begin{bmatrix} 0& 1  \end{bmatrix},
\end{aligned}
\end{equation*}
and assume that the sensor is protected. 
Then any input influences the output which is protected, so $\delta(1)$$=$$+\infty$.
However, if $A(2,1)$$=$$0$, the transfer function from the actuator to the sensor is 0, which implies $\delta(1)$$=$$1$. 
\end{example}

Lack of robustness of $\delta$ has two consequences.
Namely, an actuator that appears to be secure in one realization of the system, may be vulnerable in another.
Thus, to find actuators that are vulnerable, one should calculate $\delta$ for different realizations of $A,B,C$.  
Due to NP--hardness, this cannot be done for large scale networked control systems. 
Additionally, even if we are able to go through all the realizations of matrices $A,B,C$ and calculate indices, ensuring that $\delta$ of every actuator is large enough for every realization may require a significant security budget. 
Naturally, we may first focus on defending those actuators that are vulnerable in any realization of the system.
However, the question to answer is if we can find these actuators efficiently.

\begin{remark}
We assume that system variations occur infrequently compared to the time scale of the perfectly undetectable attacks. Hence, to the attacker, the system is linear and time-invariant.
\end{remark}

\subsubsection{Full Model Knowledge Attacker}The third issue arises due to Assumption~\ref{assumption:attacker_full_model_knowledge}.(2).
 If the system is of a large scale, the assumption that the attacker possesses the exact knowledge of the entire realization $A,B,C$ may be unrealistic. 
Lack of the full model knowledge represents a serious disadvantage for the attacker. 
Even if the attacker's knowledge is slightly inaccurate, he/she can get detected~\cite{teixeira2012revealing}.
For this reason, {Assumption~\ref{assumption:attacker_full_model_knowledge}.(2)} can result in the index being too conservative, and lead to unnecessary spending of security budget.

 \subsubsection{Replacement of $\delta$}
Due to the aforementioned three deficiencies, $\delta$ is not practical to be used in large scale networked control systems.
Therefore, in the next section, we introduce a robust security index $\delta_r$ that is based on a structural model of the system.  
We then argue in Section~\ref{section:upper_bound_properties} that $\delta_r$ represents a good candidate for replacing $\delta$ in large scale systems.
Particularly, $\delta_r$ can be calculated efficiently and it is robust with respect to system variations. 
Furthermore, having a small value of 
$\delta_r$ indicates that an actuator is vulnerable in any realization of the system, both in respect of  the attacker with the full model knowledge and the one with limited.

\section{Robust Security Index $\delta_r$~\label{section:upper_bound}}
In this section, we introduce an upper bound on the security index $\delta$. 
Based on this bound, we define the robust security index $\delta_r$.  
Prior to that, we introduce some graph theory preliminaries and a structural model of the system. 
Proofs of the results from this section are available in Appendix~\ref{appendix:proofs_2}.

\subsection{Graph Theory}

Let $\mathcal{G}=(\mathcal{V},\mathcal{E})$ be a \textit{directed graph}, with the set of \textit{nodes} $\mathcal{V}=\{v_1,\ldots,v_n\}$, and the set of \textit{directed edges} $\mathcal{E} \subseteq \mathcal{V} \times \mathcal{V}$. 
We denote by $\mathcal{N}^{\text{in}}_{v_i}=\{v_j \in \mathcal{V}: (v_j,v_i) \in \mathcal{E} \}$ the \textit{in--neighborhood} of  $v_i$. 
We say that two nodes $v_j$ and $v_k$ are \textit{non-adjacent} if there exists no edge in between them. 
Otherwise, we say they are \textit{adjacent}. 
A \textit{directed path} from $v_{j_1}$ to $v_{j_l}$ is a sequence of nodes $v_{j_1},v_{j_2},\ldots,v_{j_l}$, where $(v_{j_k},v_{j_{k+1}}) \in \mathcal{E}$ for $1\leq k < l$.  
A directed path that does not contain repeated nodes is called a \textit{simple directed path}.  
A \textit{vertex separator} (resp. an \textit{edge separator}) of  non-adjacent nodes $v_a$ and $v_b$ is a subset of nodes $\mathcal{V}' \subseteq \mathcal{V} \setminus (v_a \cup v_b)$ (resp. edges $\mathcal{E}' \subseteq \mathcal{E}$) whose removal deletes all the directed paths from $v_a$ to $v_b$. 
If each edge $(v_i,v_j)$ is assigned with weight $w_{v_iv_j}$, the cost of edge separator $\mathcal{E}'$ is defined as $\sum_{(v_i,v_j)\in \mathcal{E}'} w_{v_iv_j}$. 

\subsection{Structural Model}

The upper bound and the robust index we introduce in this section are based on a structural model $[A],[B],[C]$ of the system~\cite{dion2003generic}. 
The structural matrix $[A]\in \mathbb{R}^{n_x \times n_x}$ has only binary elements. 
If $[A](i,j)$$=$$0$, then $A(i,j)$$=$$0$ for every realization $A$. 
If $[A](i,j)$$=$$1$, $A(i,j)$ can take any value from $\mathbb{R}$. 
Same holds for matrices $[B]\in \mathbb{R}^{n_x \times n_u}$ and $[C]\in \mathbb{R}^{(n_{y}+n_{e}) \times n_x}$. 
On the one hand, this model is less informative, since it does not use the exact values of the coefficients. 
On the other hand, this also makes it more robust to system variations, which are to be expected in large scale networked systems.

We restrict our attention to a special case of matrices $[B]$ and $[C]$. 
We assume that each actuator directly influences only one state, and each sensor measures only one state. 
These are commonly adopted simplifying assumptions in sensor and actuator placement problems for large scale networked control systems~\cite{6762966,7122316,7524914}. 
Additionally, to ensure that every $B$ has a full column rank, we assume that $[B]$ has a full column rank and exclude realizations of $[B]$ where an actuator is idle (it does not influence any state). 

\begin{assumption} \label{assumption:BandCmatrix}
Let $e_i$ be the $i$-th vector of the canonical basis of appropriate size.
We assume: 
(1)~$[B]$$=$$[e_{i_1}\ldots e_{i_{n_u}}]$;
(2)~$[B]$ has a full column rank;
(3) If $[B](i,j)$$=1$, then $B(i,j)$$\neq0$ for every realization $B$;    
(4)~$[C]=[e_{j_1} \ldots e_{j_{n_y+n_e}} ]^T$. 
\end{assumption}

Assumptions~\ref{assumption:BandCmatrix}.(1)--\ref{assumption:BandCmatrix}.(3) are necessary for derivation of the results that follow. 
Assumption~\ref{assumption:BandCmatrix}.(4) is introduced to simplify the presentation, and the results can be generalized to the case when this assumption does not hold.  

We now introduce a graph $\mathcal{G}=\{\mathcal{V},\mathcal{E}\}$ of the structural model $[A],[B],[C]$. 
The set of nodes is $\mathcal{V}=\mathcal{X} \cup \mathcal{U} \cup \mathcal{Y}$, where 
$\mathcal{X}=\{x_1,\ldots,x_{n_x}\}$ is the set of states, $\mathcal{U}=\{u_1,\ldots,u_{n_u}\}$ is the set of actuators, and $\mathcal{Y}=\{y_1,\ldots,y_{n_y+n_e}\}$ is  the set of sensors.
\footnote{In the remainder of the paper, we substitute the joint set of components $\mathcal{I}$ with the sets of actuators $\mathcal{U}$ and sensors $\mathcal{Y}$. }
The set of edges is $\mathcal{E}=\mathcal{E}_{ux} \cup \mathcal{E}_{xx} \cup \mathcal{E}_{xy}$, where $\mathcal{E}_{ux}=\{(u_j,x_i): [B](i,j) \neq 0\}$ are the edges from the actuators to the states, 
$\mathcal{E}_{xx}=\{(x_j,x_i):[A](i,j) \neq 0\}$ are the edges in between the states, and $\mathcal{E}_{xy}=\{(x_j,y_i): [C](i,j) \neq 0\}$ are the edges from the states to the sensors. 
The extended graph is given by $\mathcal{G}_t=\{\mathcal{V}\cup t,\mathcal{E}_t\}$, where $\mathcal{E}_t = \mathcal{E} \cup \{(y_i,t): \forall y_i \in \mathcal{Y}\}$. 
In what follows, we use $\mathcal{G}_t$ to derive an upper bound  on $\delta$. 
We first clarify how this graph is constructed on an example. 

\begin{figure}[t]
    \centering
  \includegraphics[width=80mm]{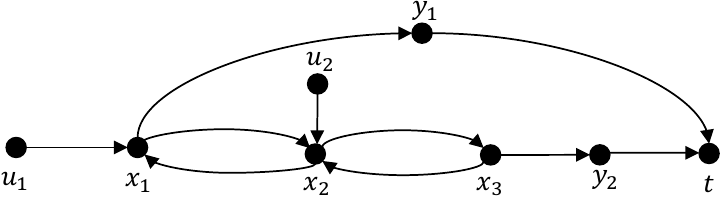}
  \caption{The extended graph $\mathcal{G}_t$ (Example~\ref{example:graph_example_1}). }
  \label{figure:graph_new_graph1}
\end{figure}

\begin{example} \label{example:graph_example_1}
Let the structural matrices be given by
\begin{align*}
[A]=\begin{bmatrix} 0  & 1 & 0 \\ 1 & 0 & 1 \\ 0 & 1 & 0 \end{bmatrix}, \hspace{2mm}[B]=\begin{bmatrix} 1 & 0 \\ 0 & 1\\ 0 & 0 \end{bmatrix},  \hspace{2mm}
 [C]=\begin{bmatrix} 1& 0 & 0 \\0& 0 & 1 \end{bmatrix}.
 \end{align*}
The extended graph $\mathcal{G}_t$ is shown in Fig.~\ref{figure:graph_new_graph1}.
\end{example}

\subsection{Upper Bound on $\delta$}

We now introduce Theorem~\ref{proposition:sec_index_sufficient_condition}, where we derive an upper bound on $\delta$ using $\mathcal{G}_t$ and vertex separators. 
Theorem~\ref{proposition:sec_index_sufficient_condition} is inspired by~\cite{7479526}, where the connection between the existence of perfectly undetectable attacks and the size of the minimum vertex separator was introduced.

\begin{theorem} \label{proposition:sec_index_sufficient_condition} 
Let $\mathcal{G}_t$ be the extended graph, $\mathcal{U}_a$ and $\mathcal{Y}_a$ be the attacked actuators and sensors, respectively, $u_i $$\in$$\mathcal{U}_a $, and
\begin{equation}\label{eqn:set_xa}
\mathcal{X}_a=\{ x_j \in \mathcal{X}:(u_k,x_j) \in \mathcal{E}_{ux},  u_k \in \mathcal{U}_a \setminus u_i \}.
\end{equation}
If $\mathcal{X}_a \cup \mathcal{Y}_a$ is a vertex separator of $u_i$ and $t$ in graph $\mathcal{G}_t$, then  
$\delta(u_i) \leq |\mathcal{U}_a|+|\mathcal{Y}_a|$
for any realization of matrices $A,B,C$. 
\end{theorem}

The intuition behind Theorem~\ref{proposition:sec_index_sufficient_condition} is the following. 
An attack against $u_i$ can be thought of as the attacker injecting a flow into the network through $u_i$. 
To stay perfectly undetectable, he/she wants to prevent the flow reaching the operator modeled by $t$. 
The attacker uses a simple strategy where he/she injects negative flows into the states $\mathcal{X}_a$ using the actuators $\mathcal{U}_a\setminus u_i$, and cancels out the flows going through these states. 
The same strategy is applied in the case of $\mathcal{Y}_a$. 
If $\mathcal{X}_a $$\cup \mathcal{Y}_a$ is a vertex separator of $u_i$ and $t$, then the  flow is successfully canceled out, so the attack remains perfectly undetectable. 
Furthermore, this strategy can be applied for any realization $A,B,C$. 

\begin{example} 
Let $\mathcal{G}_t$ be as shown in Fig.~\ref{figure:graph_new_graph1}, $u_1$ be the actuator for which we are calculating the upper bound, and assume $\mathcal{U}_a$$=$$\{u_1,u_2\}$ and $\mathcal{Y}_a$$=$$\{y_1\}$.  
Then $\mathcal{X}_a$$=$$\{x_2\}$. One can notice that by removing $\mathcal{X}_a \cup \mathcal{Y}_a$$=$$\{x_2,y_1\}$, we delete all the directed paths from $u_1$ to $t$. 
Thus, $\mathcal{X}_a$$ \cup$$ \mathcal{Y}_a$ is a vertex separator of $u_1$ and $t$, so $\delta(u_1)$$\leq $$|\mathcal{U}_a|$$+$$|\mathcal{Y}_a|$$=$$3$ in any realization of the system.  
\end{example}

\subsection{Robust Security Index $\delta_r$: Problem Formulation}
We now use  Theorem~\ref{proposition:sec_index_sufficient_condition} to introduce the robust security index $\delta_r(u_i)$ for every $u_i \in \mathcal{U}$. 
Essentially, $\delta_r(u_i)$ is the best possible upper bound from Theorem~\ref{proposition:sec_index_sufficient_condition}. 

\begin{problem} \label{problem:sec_index_upper_bound} \textit{Calculating $\delta_r$}
\begin{equation*}
\begin{aligned}
&\underset{\mathcal{U}_a , \mathcal{Y}_a}{\text{minimize } } \hspace{1mm}\delta_r(u_i)=|\mathcal{U}_a|+|\mathcal{Y}_a| \\
	&\text{subject to }\mathcal{X}_a \text{ is given by}~\eqref{eqn:set_xa}, \hspace{-2mm}&&\text{(C1)}\\
		&\hspace{15mm } \mathcal{Y}_a \text{ are not protected}, &&\text{(C2)}\\
			&\hspace{15mm } \mathcal{X}_a \cup \mathcal{Y}_a \text{ is a vertex separator of }u_i \text{ and }  t, &&\text{(C3)}\\
		&\hspace{15mm } u_i \in \mathcal{U}_a. &&\text{(C4)} 
\end{aligned}
\end{equation*}
\end{problem}
The objective reflects our goal to find an upper bound with the smallest possible value.
The constraints: 
(C1)  and (C2) ensure that the separator consists only of the states $\mathcal{X}_a$ for which there exists an actuator from $\mathcal{U}_a \setminus u_i$ adjacent to them, and unprotected sensors $\mathcal{Y}_a$; 
(C3) ensures that $\mathcal{X}_a \cup \mathcal{Y}_a$ is a vertex separator of $u_i$ and  $t$;
(C4) ensures that $u_i$ is included in the attacked components.

\begin{remark}
Just as Problem~\ref{problem:sec_index_perfectly_undetectable}, Problem~\ref{problem:sec_index_upper_bound}  does not 
have to be solvable. 
This occurs when there exists a directed path in between $u_i$ and a protected measurement, which cannot be intersected by a vertex separator. 
In that case, the attacker cannot in general use the previously introduced strategy, so we adopt $\delta_r(u_i)=+\infty$. 
Additional interpretations of $\delta_r$ being equal to $+\infty$ are provided in Section~\ref{section:upper_bound_properties}. 
\end{remark}
\begin{remark}
In the structural systems theory, it is common to use the structural model to derive results that hold for almost any realization of the system~\cite{dion2003generic}.
We depart from this type of analysis, that is, the robust security index $\delta_r$ is in general not equal to $\delta$
in almost any realization (see Section VII). 
\end{remark}

In the next section, we argue that $\delta_r$ is a good candidate to replace $\delta$ in large scale systems. 
Particularly, we show that $\delta_r$ can be efficiently calculated by solving the minimum $s$--$t$ cut problem.
Additionally, the fact that $\delta_r$ is derived based on the structural model of the system makes  it robust to system variations. 
Finally, $\delta_r$ can also be related to different types of limited model knowledge attackers.

\section{Properties of $\delta_r$~\label{section:upper_bound_properties}}

We now outline properties of $\delta_r$.
Before we move to the analysis, we revisit the minimum $s$--$t$ cut problem. 
Proofs of the results from this section are available in Appendix~\ref{appendix:proofs_3}. 

\subsection{Minimum $s$--$t$ Cut Problem}

Let $\mathcal{G}(\mathcal{V},\mathcal{E})$ be a directed graph, the source $s$ and the sink~$t$ be the elements of $\mathcal{V}$, and assume that weight $w_{v_iv_j}$ is associated to each edge $(v_i,v_j) \in \mathcal{E}$. 
A partition of $\mathcal{V}$ into $\mathcal{V}_s$ and $\mathcal{V}_t=\mathcal{V} \setminus \mathcal{V}_s$, such that $s \in \mathcal{V}_s$ and $t \in \mathcal{V}_t$, is called an $s$--$t$ cut. We define the cut capacity as 
$$ C(\mathcal{V}_s) = \sum_{\{(v_i,v_j)\in \mathcal{E}: v_i \in \mathcal{V}_s, v_j\in \mathcal{V}_t \}} w_{v_iv_j}.$$
The minimum cut problem can then be formulated as

\begin{equation}\label{prob:mincut}
\underset{\mathcal{V}_s}{\text{minimize} } \hspace{1mm}C(\mathcal{V}_s) 
\hspace{5mm}\text{subject to} \hspace{2mm}\mathcal{V}_s,\mathcal{V}_t \text{ is an $s$--$t$ cut}. 
\end{equation}
The minimum $s$--$t$ cut problem can also be interpreted as the problem of finding a minimum cost edge separator of $s$ and $t$.
Once~\eqref{prob:mincut} is solved, this separator can be recovered from $\mathcal{V}_s$ as $\mathcal{E}_c=\{(v_i,v_j)\in \mathcal{E}: v_i \in \mathcal{V}_s, v_j\in \mathcal{V}_t \}$, and its cost is $C(\mathcal{V}_s)$.

\subsection{ Efficient Computation}
In contrast to $\delta$ that is NP--hard to calculate, the exact value of $\delta_r$ can be obtained efficiently.
Particularly, the optimal value of  Problem~\ref{problem:sec_index_upper_bound} can be calculated by solving the minimum $s$--$t$ cut problem (Proposition~\ref{theorem:security_index_ts_cut}), which can  be done in polynomial time using well established algorithms such as~\cite{stoer1997simple}. 
We remark that Proposition~\ref{theorem:security_index_ts_cut} extends the previous findings on the static security index~\cite{6160456}, 
where an upper bound was also obtained by solving the minimum $s$--$t$ cut problem.

The first step towards proving Proposition~\ref{theorem:security_index_ts_cut} is to transform $\mathcal{G}_t$ to a convenient graph $\mathcal{G}_{u_i}=(\mathcal{V}_{u_i},\mathcal{E}_{u_i})$, with an additional set of edge weights $\mathcal{W}_{u_i}$. 
This graph is dependent on actuator $u_i$ for which we are calculating $\delta_r(u_i)$. 
In what follows, we explain how $\mathcal{G}_{u_i}$ is constructed. 
We use the following terminology: $x_j \in \mathcal{X}$ is said to be of Type~1, if it is adjacent to $u_k \in \mathcal{U}\setminus u_i$. Otherwise, $x_j$ is of Type~2.

\begin{remark}
In~\cite{7479526}, it was explained how to construct a graph for finding a minimum vertex separator. 
However, in our case, not all the states can be removed, and protected sensors are possible, so the graph needs to be adjusted accordingly. 
\end{remark}

The set $\mathcal{V}_{u_i}$ contains  the following nodes:
(1)~$u_i$ and $t$ (the source and the sink node);
(2)~$x_{j_{in}}$ and $x_{j_{out}}$ for every $x_j$ of Type~1;
(3)~Every $x_j$ of Type 2.
The sets $\mathcal{E}_{u_i}$ and $\mathcal{W}_{u_i}$ are constructed according to the following rules.
\begin{itemize} 
\item[(1)]  If $(u_i,x_j) \in \mathcal{E}_{ux}$, then  $(u_i,x_j) \in \mathcal{E}_{u_i}$ and $w_{u_ix_j }=+\infty$.
\item[(2)] For every $(x_{j},x_{k}) \in \mathcal{E}_{xx}$, $x_{j}\neq x_{k}$, we add an edge of the weight $+\infty$ to $\mathcal{E}_{u_i}$ subject to the following rules:
\\ - If $x_j$  is Type~1 and $x_k$ is Type~1, $(x_{j_{out}},x_{k_{in}}) \in \mathcal{E}_{u_i}$;
\\ - If $x_{j}$ is Type~1 and $x_{k}$ is Type~2, $(x_{j_{out}},x_{k}) \in \mathcal{E}_{u_i}$;
\\ - If $x_{j}$ is Type~2 and $x_{k}$ is Type~1, $(x_{j},x_{k_{in}}) \in \mathcal{E}_{u_i}$;
\\ - If $x_j$  is Type~2 and $x_k$ is Type~2, $(x_{j},x_{k}) \in \mathcal{E}_{u_i}$.
\item[(3)]  For every $x_{j_{in}}$ and $x_{j_{out}}$ that correspond to the state $x_j$ of Type~1, $(x_{j_{in}},x_{j_{out}}) \in \mathcal{E}_{u_i}$ and $w_{x_{j_{in}}x_{j_{out}} }=1$.  
\item[(4)] For every $x_j$ of Type~1 (resp. Type~2) that is measured, we add $(x_{j_{out}},t)$ (resp. $(x_{j},t)$) to $\mathcal{E}_{u_i}$.  If any of the sensors measuring $x_j$ is protected, we set the edge weight to~$+\infty$. Otherwise, the edge weight equals to the number of unprotected sensors measuring $x_j$. 
 \end{itemize}

  \begin{figure}[t]
    \centering
  \includegraphics{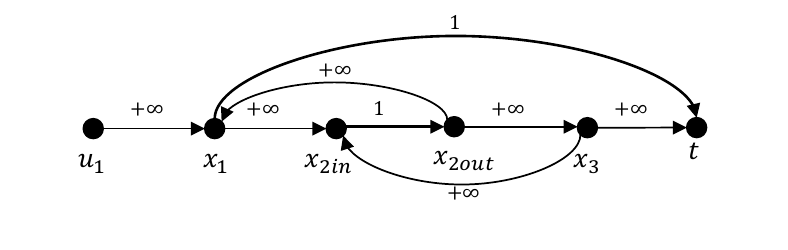}
  \caption{The graph $\mathcal{G}_{u_1}$ (Example~\ref{example:graph_example_2}).  }
  \label{figure:graph_new_graph2}
\end{figure}

\begin{example}\label{example:graph_example_2}
Assume the same structural matrices as in Example~\ref{example:graph_example_1}. Let the first sensor be unprotected, and the second one protected. 
The graph $\mathcal{G}_{u_1}$ constructed for the purpose of solving Problem~\ref{problem:sec_index_upper_bound} for actuator $u_1$ is shown in Fig.~\ref{figure:graph_new_graph2}. 
\end{example}

We now introduce Proposition~\ref{theorem:security_index_ts_cut}, which tells us that we can calculate the optimal value of Problem~\ref{problem:sec_index_upper_bound} by solving the minimum $u_i$--$t$ cut problem in $\mathcal{G}_{u_i}$.

\begin{proposition} \label{theorem:security_index_ts_cut}
 Let $\delta_r(u_i)$ be the optimal value of Problem~\ref{problem:sec_index_upper_bound}, and ${\delta}^*$ be the optimal value of the minimum $u_i$--$t$ cut problem in $\mathcal{G}_{u_i}$. 
If Problem~\ref{problem:sec_index_upper_bound} is  solvable, then $\delta_r(u_i)$$=$${\delta}^*+1$ (Statement~1). 
Otherwise, $\delta_r(u_i)$$=$${\delta}^*$$=$$+\infty$ (Statement~2). 
\end{proposition}

\subsection{ Robustness}
The second important property of $\delta_r$ is its robustness to system variations. 
Mainly, $\delta_r$ is derived based on the structural model $[A],[B],[C]$, which does not use the exact values of the system parameters. 
Hence, $\delta_r$ has the same value for any realization $A,B,C$, which is not the case with $\delta$.

\subsection{Relation of $\delta_r$ to Different Types of Attackers}
We now explain how $\delta_r$ is related to the full model knowledge attacker.
We also introduce two new attacker types without the full model knowledge, and discuss their relation to $\delta_r$. 
To distinguish between the different attacker types, in the remainder of the paper we refer to the full model knowledge attacker as the Type~1 attacker, and to the new attackers as the Type~2 and the Type~3 attackers.

\subsubsection{Type~1 Attacker}
Recall that $\delta_r(u_i)$$\geq$$ \delta(u_i)$ holds for any $u_i$$ \in $$\mathcal{U}$ and any realization $A,$$B,$$C$. 
Thus, a small $\delta_r(u_i)$ indicates serious vulnerability in respect of the Type~1 attacker. 
Particularly, not just that this attacker can conduct a perfectly undetectable attack against $u_i$ using a small number of components, but he/she can do that in any realization.

Unfortunately, as it will be shown in Section~\ref{section:simulations}, $\delta_r(u_i)$ is not a tight upper bound of $\delta(u_i)$. 
Thus, a large $\delta_r(u_i)$ does not mean that $u_i$ is secured from the Type~1 attacker.
For instance, although a solution of Problem~\ref{problem:sec_index_upper_bound} is $\mathcal{U}_a$$\cup $$\mathcal{Y}_a=\{u_i,y_{j},y_k\}$, it may exist a realization in which $u_i$ and $y_j$ are sufficient to conduct a perfectly undetectable attack against $u_i$.
However, the Type~1 attacker then needs to be sure that this realization is present.
If the realization occurs rarely, the attacker may need to wait for a long time, which increases his/her chances to be discovered in between. 
To avoid this, the Type~1 attacker may want to compromise the sensors and actuators which would allow him/her to conduct a perfectly undetectable against $u_i$ for any realization $A,B,C$.
Interestingly, the minimum number of sensors and actuators that enables this is $\delta_r(u_i)$.

\begin{proposition} \label{prop:FMNA_resources}
Let  $\mathcal{U}_a$ and $\mathcal{Y}_a$ be attacked actuators and sensors, respectively. If the Type~1 attacker can conduct a perfectly undetectable that actively uses $u_i$$\in$$\mathcal{U}_a$ for any realization of $[A],[B],[C]$, then $|\mathcal{U}_a|$$+$$|\mathcal{Y}_a|$$\geq $$\delta_r(u_i)$ must hold. 
\end{proposition}

Proposition~\ref{prop:FMNA_resources} tells us that having large $\delta_r(u_i)$ prevents the Type~1 attacker to easily gather resources that allow him/her to attack $u_i$ in any system realization.
The following corollary directly follows from the proof of Proposition~\ref{prop:FMNA_resources}. 

\begin{corollary}
If $\delta_r(u_i)=+\infty$, then there exist realizations of $[A],[B],[C]$ in which $\delta(u_i)=+\infty$. 
\end{corollary}

\subsubsection{Type~2 Attacker}
We now show that having $\delta_r(u_i)$ small implies that $u_i$ is vulnerable even if the attacker does not know the entire realization $A,B,C$. 
Particularly, we introduce the Type~2 attacker with resources limited to a local model knowledge and measurements. 
We then prove that if this attacker compromises the right combination of $\delta_r(u_i)$ components, he/she can attack $u_i$ and remain perfectly undetectable.

 \begin{assumption}\label{assumption:attacker_knowledge_2}
The Type~2 attacker: 
(1)~Can read and change the values for attacked control signals $\mathcal{U}_a$ and measurements $\mathcal{Y}_a$ arbitrarily; 
(2)~Possesses the knowledge of $[A],[B],[C]$ and of the rows $A(j,:)$, $B(j,:)$ that correspond to every state $x_j$ that is adjacent to an actuator from $\mathcal{U}_a$; 
(3)~Knows for every~$k$: $x_j(k)$ for any $x_j$ that is adjacent to an actuator from $\mathcal{U}_a$, and  
$x_l(k)$ for any $x_l \in \mathcal{N}^{\text{in}}_{x_j}$;
(4)~Wants to ensure an attack remains perfectly undetectable.
\end{assumption}

The Type~2 attacker's knowledge is limited to the structural model and the rows of $A$ and $B$ that correspond to actuators~$\mathcal{U}_a$. 
Thus, this attacker does now know the entire realization $A,B,C$.  
The attacker is also assumed to know the values of the states adjacent to $\mathcal{U}_a$ and their in-neighbors. 
The attacker can obtain these values by deploying additional sensors, but can also get this information for free.  
Namely, control algorithms sometimes base decision on the neighboring and local state to achieve better performance~\cite{7940093}. 
Hence, if the attacker remains undetected, nodes may continue sending the state information to the compromised actuators, not knowing that these actuators are controlled by the attacker.

Proposition~\ref{theorem:restricted_attacker} that we introduce next relates the Type~2 attacker to $\delta_r$.
Before we proceed to the proposition, we point out that the assumption $x(0)$$=$$0$, $u$$=$$0$ is not without loss of generality for this result to hold, as explained later.

\begin{proposition} \label{theorem:restricted_attacker}
Let $\mathcal{U}_a$ and $\mathcal{Y}_a$ be attacked actuators and sensors, respectively, $u_i \in \mathcal{U}_a $, and $\mathcal{X}_a$ be defined as in~\eqref{eqn:set_xa}. 
The Type~2 attacker can conduct a perfectly undetectable attack in which $u_i$ is actively used in any realization of $[A],[B],[C]$ if and only if $\mathcal{X}_a \cup \mathcal{Y}_a$ is a vertex separator of $u_i$ and $t$ in $\mathcal{G}_t$. 
\end{proposition}

The result has two consequences. 
Firstly, recall that $\delta_r(u_i)$ equals the minimum number of components that ensures $\mathcal{X}_a $$\cup$$ \mathcal{Y}_a$ is a vertex separator of $u_i$~and~$t$, with $u_i $$\in$$ \mathcal{U}_a$.
This implies that the Type~2 attacker with the right combination of  $\delta_r(u_i)$ components can conduct a perfectly undetectable attack against $u_i$ in any system realization. 
Particularly, it follows from the proof that the Type~2 attacker can then use a strategy similar to the one introduced to prove Theorem~\ref{proposition:sec_index_sufficient_condition}.
Yet, the strategy is implemented on-line and in a feedback fashion, based on the knowledge of local states and measurements.
This is the reason why a steady state assumption is required.
For instance, if $u$ starts changing during the attack, the Type~2 attacker can be revealed (see  Section~\ref{section:simulations}).
Secondly, same as for the Type~1 attacker,  $\delta_r(u_i)$ is the minimum number of components that allows the Type~2 attacker to conduct a perfectly undetectable attack against $u_i$ in any system realization. 
Overall, a small $\delta_r(u_i)$ implies $u_i$ is vulnerable even though the attacker does not posses the full model knowledge.

\subsubsection{Type~3 Attacker}
While the previous two propositions show that a small value of $\delta_r(u_i)$ implies that $u_i$ is vulnerable, a perhaps more interesting question to answer is if a large $\delta_r(u_i)$ implies that $u_i$ is secured.
Unfortunately, we cannot make such a claim.
Namely, both the Type~1 and the Type~2 attackers may be able to conduct a perfectly undetectable attack against $u_i$ with less than $\delta_r(u_i)$ components in some realizations.
However, we do argue that having a large value of $\delta_r(u_i)$ provides a reasonable level of security. 
Intuitively, having large $\delta_r(u_i)$ implies that an attack against $u_i$ can trigger a large number of sensors.
To avoid being detected from these sensors,
an attacker should make a synchronized attack using other
components to cancel out the effect of the attack. Thus, the
attacker should then either ensure he/she has a very precise
model and use other actuators to cancel the effect of the attack,
or he/she needs to compromise a large number of sensors. To
illustrate this point, we introduce the Type~3 attacker.

 \begin{assumption}\label{assumption:attacker_knowledge_3}
The Type~3 attacker: 
(1)~Can read and change the values for attacked control signals $\mathcal{U}_a$ and measurements $\mathcal{Y}_a$ arbitrarily; 
(2)~Knows $[A],[B],[C]$;
(3)~Wants to ensure an attack remains perfectly undetectable.
\end{assumption}

The Type~3 attacker knows only $[A]$,$[B]$,$[C]$. 
Hence, this attacker cannot constructively use other actuators to cover an attack against $u_i$, since he/she does not know which attack signals to inject in these actuators. 
However, if the system is in a steady state, the Type~3 attacker can use the Replay attack strategy~\cite{mo2015physical} to conduct a perfectly undetectable attack against  $u_i$.
In this strategy, the attacker covers an attack against $u_i$ by compromising sufficiently many sensors, and replicating steady state values from these sensors. 
Proposition~\ref{prop:WMNA_resources} establishes the connection between the number of sensors the Type~3  attacker needs to compromise and $\delta_r(u_i)$.

\begin{proposition} \label{prop:WMNA_resources}
Let $u_i$ and $\mathcal{Y}_a$ be the attacked actuator and sensors, respectively. 
If the Type~3 attacker can attack $u_i$ and ensure the attack remains perfectly undetectable, then $|\mathcal{Y}_a|$$\geq$$\delta_r(u_i)$$-1$. 
If $\delta_r(u_i)$$=$$+\infty$, the Type~3 attacker cannot attack $u_i$ and ensure the attack remains perfectly undetectable.
\end{proposition}

In other words, if the Type~3 attacker wants to ensure the attack against $u_i$ remains perfectly undetectable, then he/she needs to compromise at least $\delta_r(u_i)-1$ sensors.
Thus, a large value of $\delta_r(u_i)$ makes an attack against $u_i$ more difficult, and the Type~3 attacker is expected to avoid such actuators.
We clarify the result further in the following example. 

\begin{example}\label{example:replay}
Let the structural matrices be given by
\begin{equation*}
[A]=\begin{bmatrix}0 &  0 \\ 1 & 1  \end{bmatrix}, \hspace{5mm} [B]=\begin{bmatrix} 1 \\ 0 \end{bmatrix}, \hspace{5mm}[C]=\begin{bmatrix} 0& 1  \end{bmatrix},
\end{equation*}
and assume the Type~3 attacker only controls actuator $u_1$. 
It can be verified that the robust security index of this actuator is $\delta_r(u_1)$$=$$2$.  
Thus, according to Proposition~\ref{prop:WMNA_resources}, the attacker needs to compromise at least $\delta_r(u_1)$$-$$1$$=$$1$ sensor to ensure that an attack against  $u_1$ remains perfectly undetectable. 
Indeed, let the realization of the system be
\begin{equation*}
\begin{aligned}
A=\begin{bmatrix}0 &  0 \\ a_{21} & a_{22}  \end{bmatrix}, \hspace{5mm} B=\begin{bmatrix} 1 \\ 0 \end{bmatrix}, \hspace{5mm}C=\begin{bmatrix} 0& 1  \end{bmatrix}.
\end{aligned}
\end{equation*}
If $a_{21}$$\neq$$0$, any attack against $u_1$ is visible in the sensor measurement. 
Since the Type~3 attacker knows only the structural model of the system, he/she does not know the exact value of  $a_{21}$.
Thus, he/she needs to compromise the sensor to ensure an attack against $u_1$ remains perfectly undetectable. 
\end{example}

\subsubsection{Summary}
The main conclusions of this subsection are as follows. 
Firstly, a small $\delta_r(u_i)$ indicates that $u_i$ is vulnerable with respect to the Type~1 and the Type~2 attackers in any realization of the system. 
Secondly, a large $\delta_r(u_i)$ does not indicate security with respect to these attackers, but it does prevent them from easily gathering resources for attacking $u_i$ in any realization of the system.
Finally, a large $\delta_r(u_i)$ indicates security  with respect to the Type~3 attacker. 
For these reasons, it is useful to derive strategies for increasing $\delta_r$. 
We consider this problem in the next section.

\section{Sensor Placement for Increasing~$\delta_r$~\label{section:improving_upper_bound}}

In this section, we discus how $\delta_r$ can be increased by placing additional sensors. 
We derive sets of suitable positions to place sensors, and then introduce two sensor placement problems with the objective to increase the robust indices of actuators. 
We show that these problems have convenient submodular structures, which allow us to efficiently obtain suboptimal solutions of these problems with guarantees on performance.
Before we move to the analysis, we introduce a necessary background on submodular optimization. 
Proofs of the results from this section are available in Appendix~\ref{appendix:proofs_4}. 

\subsection{Submodular Optimization}

We begin by introducing the definitions of submodular and nondecreasing set functions, and recalling some well known properties of these functions~\cite{krause2014submodular}.

\begin{definition} \label{definition:submodular}
Let $\mathcal{X}$$=$$\{x_1,\ldots,x_n\}$ be a finite non-empty set and  
$F$$:$$2^\mathcal{X} $$\rightarrow$$ \mathbb{R}$ be a set function. We say that $F$ is \textit{submodular} if $F( \mathcal{X}_a \cup x)$$-$$F(\mathcal{X}_a) $$\geq$$ F(\mathcal{X}_b\cup x)$$-$$F( \mathcal{X}_b)$
holds for all $\mathcal{X}_a$$\subseteq$$ \mathcal{X}_b$$\subseteq$$ \mathcal{X}$ and $x $$\in $$ \mathcal{X}$$\setminus$$\mathcal{X}_b$. We say that $F$ is \textit{nondecreasing} if $F( \mathcal{X}_a)\leq F(\mathcal{X}_b)$ holds for all $\mathcal{X}_a$$\subseteq$$ \mathcal{X}_b$$\subseteq \mathcal{X}$.

\end{definition}

\begin{lemma} \label{lemma:properties_submodular}
Let $F_1,\ldots,F_n$ be submodular and non-decreasing set functions and $c$ be an arbitrary constant. Then
$g_1(\mathcal{X}_a)=\sum_{i=1}^n F_i(\mathcal{X}_a)$ and $g_2(\mathcal{X}_a)=\min\{F_i(\mathcal{X}_a),c\}$ are submodular and nondecreasing set functions.
\end{lemma}

Submodularity has an important role in combinatorial optimization. 
Particularly, many interesting problems with submodular structure can be approximately solved in polynomial time with guarantees on performance~\cite{bach2013learning}.
In this work, we are interested in the following two problems
\begin{align} \label{eqn:sub_problem_2}
&\underset{  \mathcal{X}_p }{\text{minimize}} \hspace{2mm} |  \mathcal{X}_p| \hspace{5mm} &&\text{subject to} \hspace{2mm} F(  \mathcal{X}_p) \geq F_{\max}, \\
&\underset{ \mathcal{X}_p}{\text{maximize}} \hspace{2mm} F'( \mathcal{X}_p) \hspace{5mm} &&\text{subject to} \hspace{2mm} |  \mathcal{X}_p| \leq k_{\max}, 
\label{eqn:sub_problem_1}
\end{align}
where $F$ and $F'$ are nondecreasing and submodular set functions that satisfy $F(\emptyset)$$=$$F'(\emptyset)$$=$$0$, $F_{\max} $$\in $$\mathbb{Z}_{\geq 0}$, and $k_{\max}$$ \in $$\mathbb{Z}_{\geq 0}$.
Additionally, $F$ is assumed to be an integer valued function.
Suboptimal solutions for both~\eqref{eqn:sub_problem_2} and~\eqref{eqn:sub_problem_1} can be obtained in polynomial time with relatively good performance guarantees.  

\begin{lemma}\label{lemma:logbound}\cite[Theorem~1]{wolsey1982analysis} Let $|\mathcal{X}^*|$ be the optimal value of~\eqref{eqn:sub_problem_2}, and $H(d)$$=$$\sum_{i = 1}^{d} \frac{1}{i}$. A suboptimal solution $ \mathcal{X}_{g}$ of~\eqref{eqn:sub_problem_2} that satisfies
$|\mathcal{X}_{g}| $$\leq H$$(\text{max}_{x \in \mathcal{X}}\hspace{0.5mm} F(x) )|\mathcal{X}^*|$
can be obtained in polynomial time using the algorithm given in~\cite[Section~2]{wolsey1982analysis}.
\end{lemma}

\begin{lemma}  \label{lemma:fixbound}\cite[Proposition 4.3]{nemhauser1978analysis}
Let $F^*$ be the optimal value of~\eqref{eqn:sub_problem_1}. 
A suboptimal solution $ \mathcal{X}_{g}$ of~\eqref{eqn:sub_problem_1} that satisfies
$F'( \mathcal{X}_g) $$\geq $$(1- \frac{1}{e})F^*$
 can be obtained in polynomial time using the algorithm given in~\cite[Section 4]{nemhauser1978analysis}.
\end{lemma}

We remark that the bounds introduced in Lemmas~\ref{lemma:logbound} and~\ref{lemma:fixbound}	represent the worst case performance guarantees. 
The algorithms mentioned in the lemmas can perform better in practice.

\subsection{Suitable Locations to Place Sensors}

We now introduce a suitable set of states $\mathcal{X}_{u_i}$ connected to each actuator $u_i$. 
We show that if we place a new sensor to measure any of the states from $\mathcal{X}_{u_i}$, $\delta_r(u_i)$ is guaranteed to increase. 
Moreover, if every state adjacent to an actuator is also adjacent to a sensor, then placing a new sensor to measure a state from $\mathcal{X}_{u_i}$ is the only way to increase $\delta_r(u_i)$.  

\begin{theorem}
\label{proposition:place_to_put_sensors}
Let $\mathcal{G}_t$ be the extended graph, $u_i$ be an actuator with $\delta_r(u_i)$$ \neq $$+\infty$, and $x_{k}$$ \in$$ \mathcal{X}$ be such that there exists a directed path $u_i,x_j,\ldots,x_{k}$ in which none of the states is adjacent to an actuator from $\mathcal{U} \setminus u_i$.
Let the set of all such states be denoted with $\mathcal{X}_{u_i}$. 
Assume that a new sensor $y_l$ is placed to measure an arbitrary state from $\mathcal{X}_{u_i}$, and let $\delta_r'(u_i)$ be the robust index of $u_i$ after the placement. Then:  
\begin{itemize}
\item[(1)] $\delta_r'(u_i)=+\infty$ if $y_l$ is protected;  
\item[(2)] $\delta_r'(u_i)=\delta_r(u_i)+1$ if $y_l$ is unprotected.  
\end{itemize}
Furthermore, assume that for every $x_j \in \mathcal{X}$ for which there exists $(u_k,x_j) \in  \mathcal{E}_{ux}$, there also exists $(x_j,y_p) \in  \mathcal{E}_{xy}$. 
Then $\delta_r(u_i)$ is increased if and only if a new sensor is placed to measure a state from $\mathcal{X}_{u_i}$.   
\end{theorem}

The sets  $\mathcal{X}_{u_1}$$, \ldots,$$ \mathcal{X}_{u_{n_u}}$ introduced in the previous theorem have two important properties. 
Firstly, for every $u_i \in \mathcal{U}$, $\mathcal{X}_{u_i}$ can easily be found as follows. 
We first remove from the graph $\mathcal{G}_t$ all the states that are adjacent to an actuator from $\mathcal{U}\setminus u_i$. 
In that case, the set $\mathcal{X}_{u_i}$ is the set of all the states to which $u_i$ is connected with a directed path.
We can then apply the depth first search algorithm~\cite{cormen2009introduction} to find these states. 
Secondly, these sets are not affected by the placement of new sensors. 
Thus, if we place $n$ sensors to monitor the states from $\mathcal{X}_{u_i}$, $\delta_r(u_i)$ is guaranteed to increase by $n$. 

In what follows, we use Theorem~\ref{proposition:place_to_put_sensors} to formulate two sensor placement problems.
As we shall see, suboptimal solutions with performance guarantees can be obtained efficiently for both of these problems, even in large scale networked systems.

\begin{remark}
The sensor placement problems we introduce next are developed for increasing $\delta_r$, which does not in general imply that we increase $\delta$ at the same time. 
However, the placement of new sensors cannot decrease $\delta$ (Proposition~\ref{prop:adding_sensors_delta}), so we definitely do not degrade this index. 
In fact, we illustrate in Section~\ref{section:simulations} that by increasing $\delta_r$, we often indirectly increase $\delta$. 
Future work will investigate how to preselect some of the states from the previously introduced sets, such that we know that $\delta$ is increased for at least some classes of realizations.  
\end{remark}

\subsection{Sensor Placement Problems}

\subsubsection{Placement of Unprotected Sensors} We first discus the problem of placing unprotected sensors. 
The goal is to place these sensors such as to increase $\delta_r$ for every actuator $u_i$ by at least  $k_{u_i} $$\in$$ \mathbb{Z}_{\geq 0}$. 
We assume unprotected sensors to be inexpensive, so we do not have a sharp constraint on the number of sensors we should place. 
Yet, we still want to place the minimum number of them to achieve the desired benefit. 

Let the set of sensors be $\mathcal{Y}_s=\{y_1,\ldots,y_{n_s}\}$, and $x_{y_i}$ be the state measured by $y_i$. 
For every actuator $u_i$, we define 
$g_{u_i}(\mathcal{Y}_p)= \min\{ \sum_{y_j \in \mathcal{Y}_p}|x_{y_j} \cap  \mathcal{X}_{u_i}|,k_{u_i}\},$ where $\mathcal{Y}_p \subseteq  \mathcal{Y}_s$ is the set of newly placed sensors. 
This function equals $k_{u_i}$, if at least $k_{u_i}$ sensors from $\mathcal{Y}_p$ measure the states from $ \mathcal{X}_{u_i}$.
We then have from Theorem~\ref{proposition:place_to_put_sensors} that $\delta_r(u_i)$ is increased by at least $k_{u_i}$. 
Additionally, if every state adjacent to an actuator is also adjacent to a sensor, then $\delta_r(u_i)$ is increased exactly by $k_{u_i}$.

Let $ G(\mathcal{Y}_p)= \sum_{u_i \in \mathcal{U} } g_{u_i}(\mathcal{Y}_p)$ be the total gain achieved by placement  $\mathcal{Y}_p$. 
If $G(\mathcal{Y}_p)\geq \sum_{u_i \in \mathcal{U}} {k_{u_i}}$, then the robust indices of all the actuators are increased by the desired values. 
The problem of placing unprotected sensors is then
    \begin{equation}\label{eqn:protecting_sec_index2}
\underset{ \mathcal{Y}_p \subseteq \mathcal{Y}_s}{\text{minimize}} \hspace{1mm} |\mathcal{Y}_p| 
\hspace{10mm}
\text{subject to} \hspace{2mm}G(\mathcal{Y}_p)\geq \sum_{u_i \in \mathcal{U}} {k_{u_i}}.
\end{equation}
The objective function we are minimizing is the number of deployed sensors. 
The constraint implies that we continue placing sensors until the robust indices of all the actuators are for sure increased by the desired value. 
The following proposition shows that this problem is an instance of Problem~\eqref{eqn:sub_problem_2}, so we can find a suboptimal solution for it in polynomial time with guarantees stated in Lemma~\ref{lemma:logbound}.

\begin{proposition}  \label{theorem:improving_bound_subm_2}
Problem~\eqref{eqn:protecting_sec_index2} is an instance of Problem~\eqref{eqn:sub_problem_2}.
 \end{proposition}

\subsubsection{Placement of Protected Sensors} 
One can also consider the problem of deploying protected sensors. 
One objective could be to increase $\delta_r$ to $+\infty$ for as many actuators as possible, which would prevent the Type~3 attacker of attacking these actuators.
Since protected sensors might be expensive, we assume that the operator is limited to $k_{\max}$ sensors. 

The problem can be formulated as follows.  
Let $\mathcal{X}_p \subseteq \mathcal{X}$ be the subset of states that we want to measure using the protected sensors. 
Similar to the previous placement problem, we first define the function $g'_{u_i}(\mathcal{X}_p)= \min\{|\mathcal{X}_p \cap \mathcal{X}_{u_i}|,1\}$ for each $u_i$. 
If $g'_{u_i}(\mathcal{X}_p)=1$, then there exist a protected sensor measuring a state from $\mathcal{X}_{u_i}$, and we know from Theorem~\ref{proposition:place_to_put_sensors} that $\delta_r(u_i)=+\infty$.  
Otherwise, $g'_{u_i}(\mathcal{X}_p)=0$. 

Let $\mathcal{U}_p \subseteq \mathcal{U}$ be a subset of actuators for which we want to increase the robust indices to $+\infty$.
We can then define the objective function as $ G'(\mathcal{X}_p)= \sum_{u_i \in \mathcal{U}_p} g'_{{u_i}}(\mathcal{X}_p).$
The value of this function equals the number of actuators whose robust indices are equal to $+\infty$ after placing protected sensors at locations $\mathcal{X}_p$. 
Naturally, we want to maximize this gain function, with no more than $k_{\max}$ deployed sensors. 
The problem we want to solve can then be formulated as 
\begin{equation} \label{eqn:protecting_sec_index1}
\underset{ \mathcal{X}_p \subseteq \mathcal{X}}{\text{maximize}} \hspace{1mm} G'(\mathcal{X}_p)  \hspace{10mm}
\text{subject to} \hspace{1mm}|\mathcal{X}_p| \leq k_{\max}.
\end{equation}

We now show that~\eqref{eqn:protecting_sec_index1} represents an instance of~\eqref{eqn:sub_problem_1}. 
It then follows from Lemma~\ref{lemma:fixbound} that a suboptimal solution of~\eqref{eqn:protecting_sec_index1} with $1-\frac{1}{e}$ approximation ratio can be obtained in polynomial time. 

\begin{proposition}\label{theorem:improving_bound_subm_1}
Problem~\eqref{eqn:protecting_sec_index1} is an instance of Problem~\eqref{eqn:sub_problem_1}. 
\end{proposition}

\section{Illustrative Examples \label{section:simulations}}

We now discuss the theoretical developments on illustrative numerical examples.

\subsection{Comparison of $\delta$ and $\delta_r$ }

\subsubsection{Model}  
\begin{figure}[t]
    \centering
  \includegraphics[width=80mm]{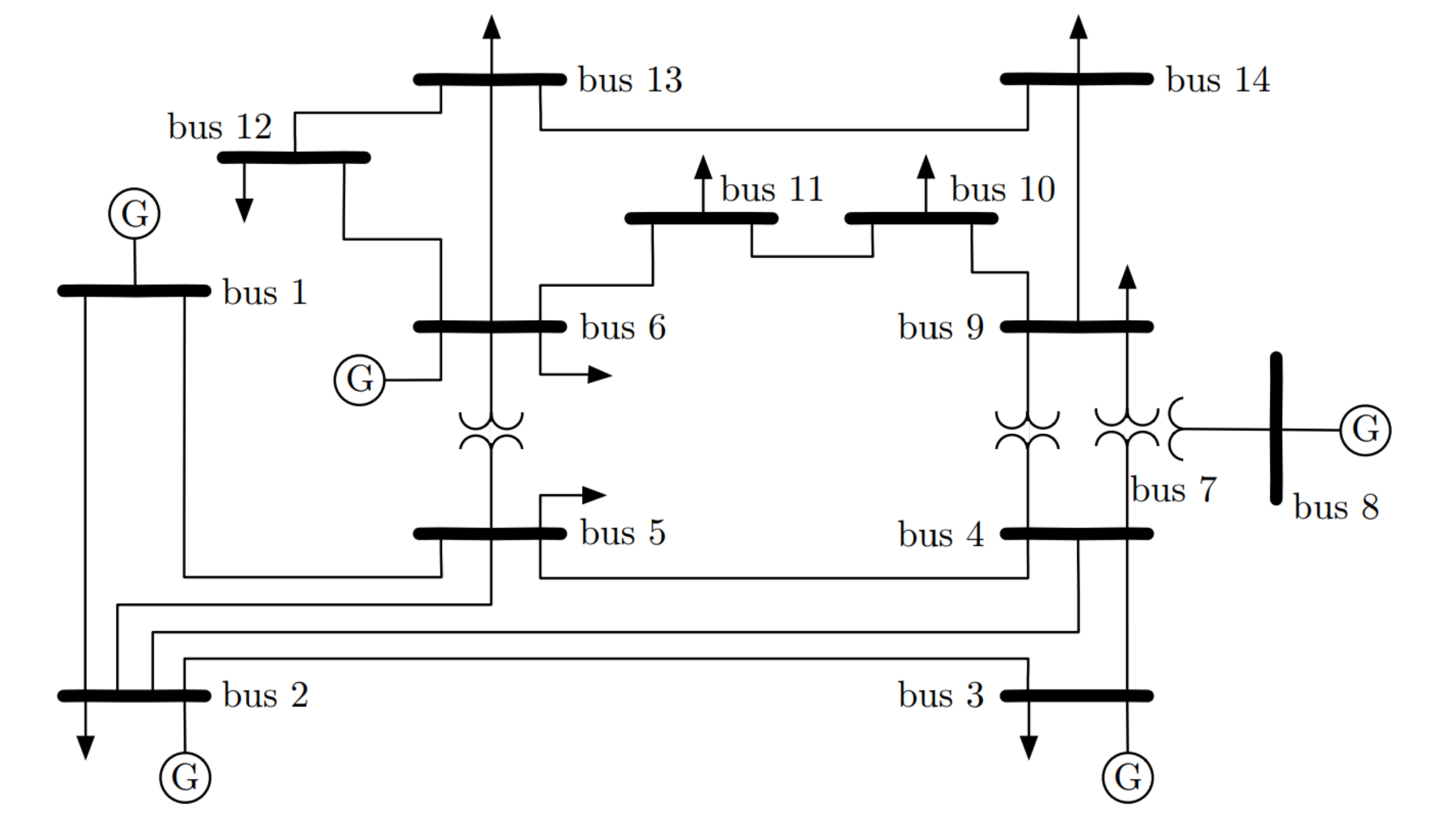}
  \caption{Schematic of IEEE 14 bus system~\cite{6545301}. }
  \label{figure:IEEE14_schematic}
\end{figure}

We consider the IEEE~14 bus system, shown in Fig.~\ref{figure:IEEE14_schematic}.  
The system is controlled 
using 5 generators located at buses 1,2,3,6, and~8. 
We modeled the system using linearized swing equations where the generators were represented by two states (rotor angle $\phi_i$ and frequency $\omega_i=\dot{\phi_i}$), and load buses with one state (voltage angle $\theta_i$)~\cite{4110445}. 
The parameters given in~\cite{kodsi2003modeling} were used.
The operator has access to phasor measurement units providing measurements of $\theta_1,\theta_3,\theta_5,\theta_7,\theta_9,\theta_{11}$, and~$\theta_{13}$.
We considered the following system realizations:
\begin{itemize} 
\item Normal operation, as shown in Fig.~\ref{figure:IEEE14_schematic} (Realization~1); \item Power line (Bus~4,Bus~7) switched--off (Realization~2);
\item Micro--grid consisting of Bus~3 and Generator~3 detaches from the grid (Realization~3); 
\item Measurement $\theta_1$ stops being available (Realization~4). 
\end{itemize}
We assumed that every generator and every measurement can be compromised by the attacker. 
Furthermore, the attacker is assumed to be able to attack the network by changing loads at some buses~\cite{5976424}. 
Particularly, the loads at buses $2,5,9$, and $14$ were assumed to have considerable effect to the network, and were modeled as additional actuators.

\subsubsection{Robustness}
\begin{figure}[t]
   \centering
  \includegraphics[width=70mm]{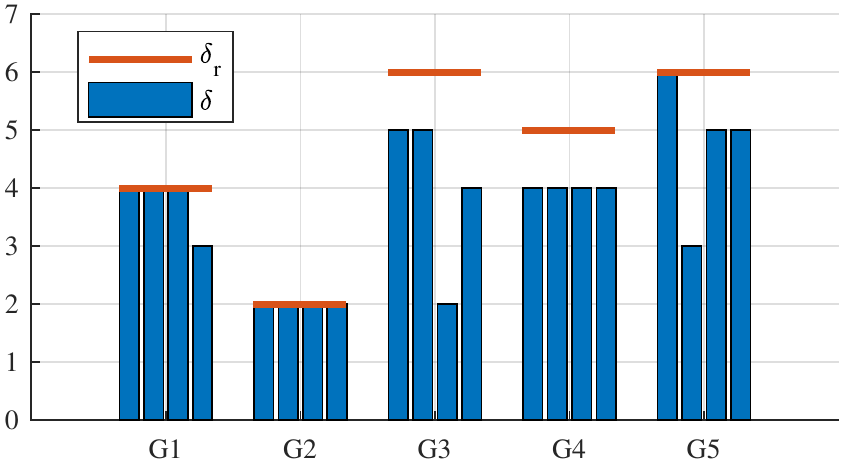}
  \caption{ The value of the security index $\delta$ and the robust security index $\delta_r$ of Generators~1--5 for different realizations of the system. }
  \label{figure:figure_sim_1}
\end{figure}

We first compare $\delta$ and $\delta_r$ in terms of robustness. 
For this purpose, we calculated the values of $\delta$ and $\delta_r$ of all the generators in the aforementioned four realizations of the system.
The results are shown in Fig.~\ref{figure:figure_sim_1}. 

Firstly, the results confirm that $\delta$ depends on realization of the system. 
Thus, if the operator decides to use $\delta$ as a security index, it is not sufficient to consider only one realization. 
For example, Generator~3 that appears to be the second most secured in Realization~1, becomes one of the two most vulnerable in Realization~3. 
A less evident observation is that the use of $\delta$ can lead to a considerable security allocation cost. 
Particularly, we see that the minimum value of $\delta$ for all the generators is quite similar (except for maybe Generator~4).
Therefore, ensuring that each generator has sufficiently large security index $\delta$ for every realization of the system may be very hard, and would require a large security investment.  

Evidently, the values of $\delta_r$ are not dependent on the realization. 
Therefore, having a small value of $\delta_r(u_i)$ implies that actuator $u_i$ is vulnerable in any system realization.  
For example, since $\delta_r(G_2)$$=$$2$,  Generator~2 can be attacked by the Type~1 and the Type~2 attackers by compromising only two components in any realization of the system. 
However, as it can be seen, $\delta_r$ is not a tight upper bound on $\delta$.
Thus, large $\delta_r$ does not necessarily imply security, which is the main drawback of $\delta_r$.
For instance, note that $\delta(G_3)$$=$$2$ in the third realization.
Hence, the Type~1 attacker can conduct a perfectly undetectable attack against Generator~3 in this realization by compromising two components, although $\delta_r(G_3)$$=$$6$.

\subsubsection{Computing $\delta$ and $\delta_r$} 
We now compare the computational efforts needed to calculate $\delta$ and $\delta_r$. 
To calculate $\delta$, we used the brute force search method explained in Section~\ref{section:properties}. 
To calculate $\delta_r$, we used the \texttt{maxflow} function that is included in Matlab R2017. 
We kept the realization of the system fixed to Realization~1, and started increasing the number of sensors by placing new sensors at random locations. 
We then measured time needed to calculate $\delta$ and $\delta_r$ for Generator $G_4$. 

\begin{figure}[t]
   \centering
  \includegraphics[width=80mm]{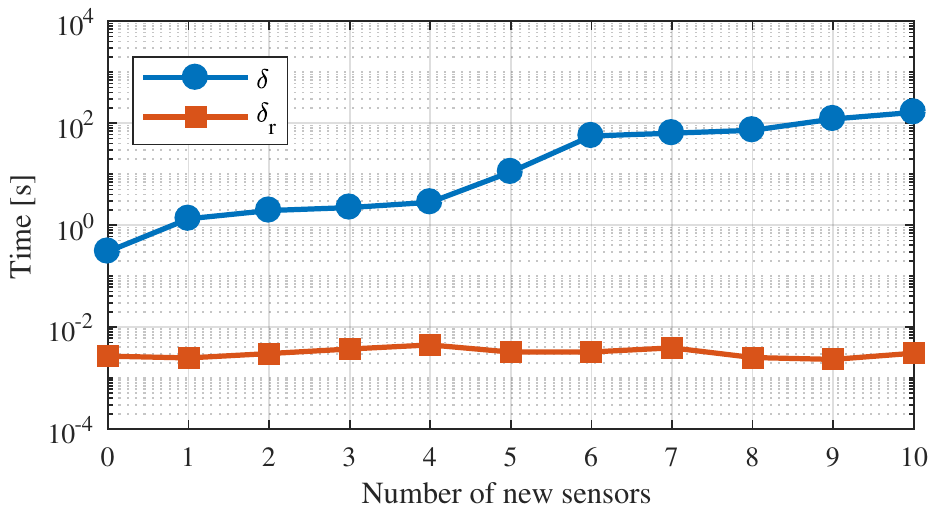}
  \caption{Computational time required for finding the exact value of $\delta$ and $\delta_r$ of Generator~4 once the number of sensors vary.  }
  \label{figure:figure_sim_2}
\end{figure}

The results are shown in Fig.~\ref{figure:figure_sim_2}. 
As expected, the effort for calculating $\delta$ grows exponentially with the number of newly added sensors.
Furthermore, note that this effort scales with the number of realization for which we want to calculate $\delta$. 
The time needed for calculating $\delta_r$ was almost not affected by placing this relatively small number of sensors, and remained below 0.01 [s] in all the cases. 
Additionally, $\delta_r$ is calculated only once, since it has the same value in any realization.   

  \begin{figure}[t]
   \centering
  \includegraphics[width=78mm]{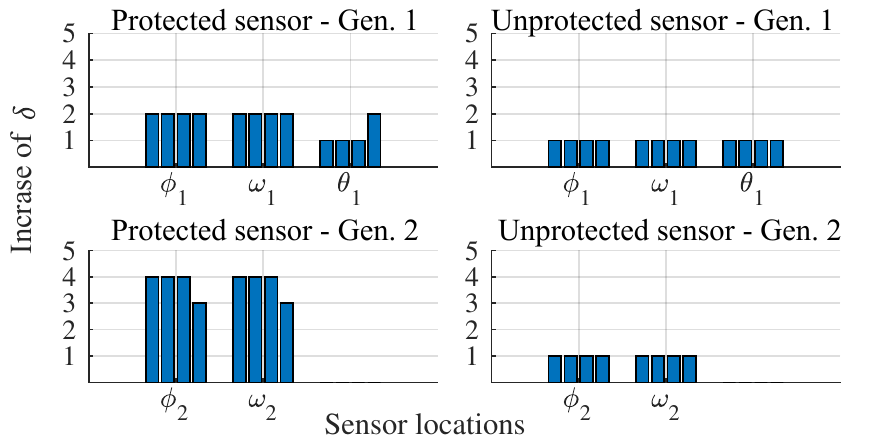}
  \caption{Increase of the security index $\delta$ for Generator~1 and Generator~2.}
  \label{figure:figure_sim_41}
\end{figure}
\subsubsection{Increasing $\delta$ and $\delta_r$}  
We now investigate if by increasing $\delta_r$ we also increase $\delta$. 
We focus on Generators~1 and~2, since these generators have the lowest values of $\delta_r$.
Using Theorem~\ref{proposition:place_to_put_sensors}, we obtained that suitable locations for placing additional sensors are $\mathcal{X}_{G_1}=\{ \phi_1,\omega_1, \theta_1\}$ for Generator~1 and $\mathcal{X}_{G_2}=\{ \phi_2,\omega_2\}$ for Generator~2.

We first investigated how the placement of one protected sensor influences $\delta$. 
We placed sensor at each of the locations from $\mathcal{X}_{G_1}$, and measured the increase of $\delta(G_1)$. 
While placing the protected sensor at these locations increases $\delta_r(G_1)$ to $+\infty$,
it can be seen from Fig.~\ref{figure:figure_sim_41} that $\delta(G_1)$ did not increase  to $+\infty$ in any of the four realizations we considered.
Yet, the increase of $\delta(G_1)$ for more than one was achieved in majority of the cases,   
which is not possible to achieve by placing an unprotected sensor (Proposition~\ref{prop:adding_sensors_delta}).
The experiment was also conducted for Generator~2.
Similarly, $\delta(G_2)$ did not increase to $+\infty$ in any of the four realizations.
However,  the placement of one protected sensor lead to increase of $\delta(G_2)$ by at least three for all the locations from $\mathcal{X}_{G_2}$ and all the realizations. 

We also considered placing one unprotected sensors at locations from $\mathcal{X}_{G_1}$, which increases $\delta_r(G_1)$ by one. 
Interestingly, from Fig.~\ref{figure:figure_sim_41}, the placement of one unprotected sensor at any of the locations from $\mathcal{X}_{G_1}$ lead to increase of $\delta(G_1)$ in all the realizations. 
The same holds for $\mathcal{X}_{G_2}$ and $\delta(G_2)$.

Overall, the experiment illustrates that by increasing $\delta_r$ we can also indirectly increase $\delta$.  
However, from the placement of protected sensors, we see that we definitely do not achieve the same level of improvement. 
This again illustrates that protecting the system against the advanced Type~1 attacker may require much more resources than protecting it against less advanced attackers such as the Type~3 attacker.

\subsection{Properties of Full and Limited Model Knowledge Attackers}

  \begin{figure}[t]
   \centering
  \includegraphics[width=40mm]{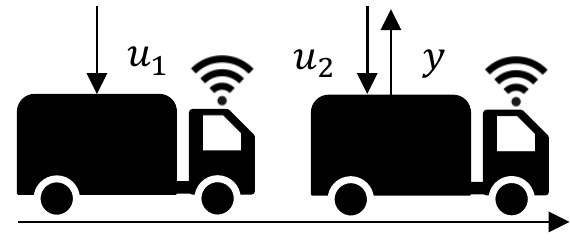}
  \caption{The platoon consisting of two autonomous vehicles. 
  Each vehicle can be controlled by the operator through the signals $u_1$ and $u_2$.
  The operator also knows the position of the second vehicle $y$. }
  \label{figure:Platon}
\end{figure}

\subsubsection{Model} 
We now illustrate  the limitations of the full and limited model knowledge attackers considered in the paper.  
For this purpose, we consider the system of two autonomous vehicles shown in Fig.~\ref{figure:Platon}.
Each vehicle is modeled by a single state representing its position relative to some moving reference frame.
The operator can control both vehicles through signals $u_1$ and $u_2$, and he/she also knows the position of the second vehicle $y=x_2$.
The operator's goal is to keep the distance between vehicles equal to $10$.  
To study this formation control problem, we use the model from~\cite{6816560} 
\begin{align*}
x(k+1)&=\begin{bmatrix}
1-2\alpha_1 & \alpha_1              \\
\alpha_2     & 1-2\alpha_2 
 \end{bmatrix}x(k)+u(k),\\
 y(k)&=\begin{bmatrix} 0 & 1\end{bmatrix} x(k),
   \end{align*}
where $\alpha_1=\alpha_2=0.1$. 
We assume that prior to the attack, $x(0)=[0\hspace{2mm}10]^T$ and $u(0)=[-1\hspace{2mm}2]^T$, so that desired behavior of the platoon is achieved.  

We consider  the Type~1 attacker and the Type~2 attacker.
Both of the attackers control $u_1$ and $y$, and have the goal to disrupt the platoon formation without the operator noticing. 
In the following, we discuss in which situations the attackers can achieve this goal. 
By $\Delta y_{F}$ (resp. $\Delta y_{L}$), we denote the difference between the measurement expected in the normal operation and the received measurement in the case of the first (resp. second) attacker. 
If the attackers are able to conduct a perfectly undetectable attack, then $\Delta y_{F}$$=$$\Delta y_{L}$$=$$0$ must hold. We also remark that the properties of the Type~2 attacker we outline next are the same as for the Type~3 attacker, so we do not explicitly consider the Type~3 attacker. 

\subsubsection{Case~1}
\begin{figure}[t]
   \centering
  \includegraphics[width=80mm]{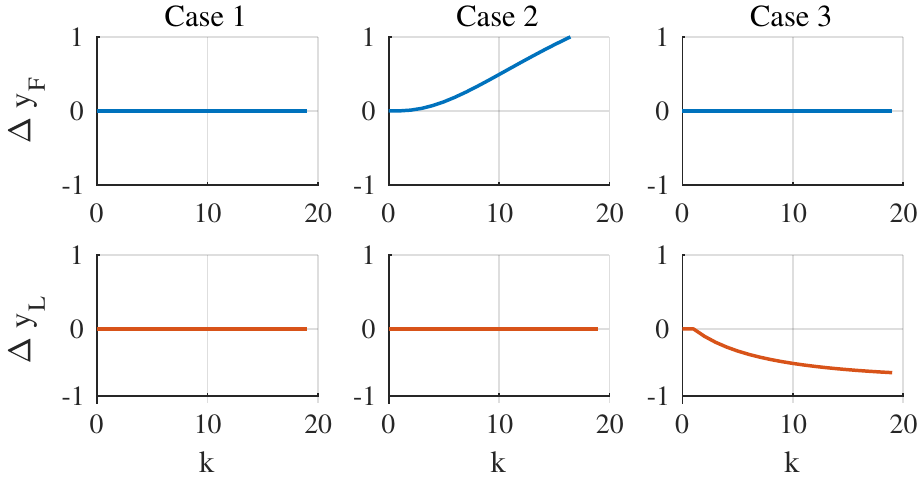}
  \caption{Differences $\Delta y_{F}$ and $\Delta y_{L}$ of the expected and attacked sensor measurements in different cases.   }
  \label{figure:figure_sim_32}
\end{figure}

The first case illustrates that both of the attackers can conduct a perfectly undetectable attack once the system is in a steady state and $u(k)=u(0)$ during the attack.  
The Type~1 attacker applies the following signals
\begin{equation}\label{eqn:full_sim}
\small
\begin{aligned}
 a^{(u_1)}_{F}(k)&= -k, \\
a^{(y)}_{F}(k+2)&=1.6a^{(y)}_{F}(k+1)-0.63 a^{(y)}_{F}(k)-0.1a^{(u_1)}_{F}(k),
\end{aligned}
\end{equation}
which is according to the strategy  introduced in the proof of Proposition~\ref{theorem:condition}. 
The Type~2 attacker applies the signals
\begin{equation}\label{eqn:lim_sim}
\small
\begin{aligned}
a^{(u_1)}_{L}(k)= -k, \hspace{5mm} a^{(y)}_{L}(k)&=-x_2(k)+y(0),
\end{aligned}
\end{equation}
which is according to the strategy introduced in the proof of Proposition~\ref{theorem:restricted_attacker}.
As we can see from Fig.~\ref{figure:figure_sim_32}, Case~1, $\Delta y_{F}$$=$$\Delta y_{L}$$=$$0$.
Hence, both of the attackers remain perfectly undetectable.
Additionally, note that in this case, the strategy~\eqref{eqn:lim_sim} reduces to the Replay attack strategy that does not require any realization knowledge. 
Thus, the Type~3 attacker that controls $u_1$ and $y$ can also use the strategy~\eqref{eqn:lim_sim}, so he/she would also remain perfectly undetectable in this case. 

\subsubsection{Case~2} 
The second case is introduced to illustrate the fragility of the Type~1 attacker with respect to modeling errors. 
Assume $u(k)$$=$$u(0)$ during the attack, and that the Type~1 attacker believes that $\alpha_2' $$=$$ 0.11$. 
He/she then applies the signals 
\begin{equation*}
\small
\begin{aligned}
 a^{(u_1)}_{F}(k)&= -k, \\
a^{(y)}_{F}(k+2)&=1.58a^{(y)}_{F}(k+1)-0.613 a^{(y)}_{F}(k)-0.11a^{(u_1)}_{F}(k).
\end{aligned}
\end{equation*}
The Type~2 attacker applies the same signals as in the previous case. 
 From Fig.~\ref{figure:figure_sim_32}, Case~2, we can see that $\Delta y_{F}$$\neq$$0$, so the Type~1 attacker is revealed. 
Since $\Delta y_{L}$$=$$0$, we see that the Type~2 attacker remains undetected. 
In general, the Type~2 attacker can also be vulnerable to modeling errors, since he/she may require precise local  model knowledge to construct the strategy in some cases.
However,  the fact that this attacker uses only a fraction of the model (in this case none), lowers his/her chances to become detected because of modeling errors. 
Evidently, the Type~3 attacker is not affected by this type of errors, since he/she knows only the system structure. 

\subsubsection{Case~3} 
Finally, assume the scenario where at $k=2$, the operator increases $u_2$ by $0.1$. 
The attackers apply the signals~\eqref{eqn:full_sim} and~\eqref{eqn:lim_sim}.
From Fig.~\ref{figure:figure_sim_32}, Case~3, we can see $\Delta y_{L}$$\neq$$0$.
This illustrates that the steady state assumption is in general required for the Type~2 attacker to remain perfectly undetectable.
The reason is that this attacker does not know neither $u_2$ nor the equation for $x_2$. 
Thus, once $y$ starts changing, the attacker cannot distinguish if this is because of the attack or a change in $u_2$. 
The same reasoning applies to the Type~3 attacker. 
We also see that $\Delta y_{F}$$=$$0$.
The reason is that the attack policy~\eqref{eqn:full_sim} can be calculated prior to the attack and implemented in a feedforward manner.
This makes the strategy completely decoupled from $x(0)$ and $u$.

\section{Conclusion and Future Work}\label  {section:conclusion}
\begin{table}
  \caption{The main properties of the attacker types we considered \newline and their relation to $\delta_r$. }
\centering
\begin{tabular}{|c|c|l|c|c|}
\hline
\multirow{2}{*}{Attacker} & \multicolumn{2}{c|}{\multirow{2}{*}{\begin{tabular}[c]{@{}c@{}}Knowledge of\\ $A,B,C$~/~$[A],[B],[C]$\end{tabular}}} & \multirow{2}{*}{\begin{tabular}[c]{@{}c@{}}Steady State\\ Assumption\end{tabular}} & \multirow{2}{*}{\begin{tabular}[c]{@{}c@{}}Relation\\  to $\delta_r$\end{tabular}} \\
                          & \multicolumn{2}{c|}{}                                                                                                  &                                                                                    &                                                                                    \\ \hline
Type~1               & \multicolumn{2}{c|}{Full~/~Full}                                                                                         & Not required                                                                       & \begin{tabular}[c]{@{}c@{}}Upper bounds\\ resources\end{tabular}                   \\ \hline
Type~2               & \multicolumn{2}{c|}{Limited~/~Full}                                                                                      & Required                                                                           & \begin{tabular}[c]{@{}c@{}}Upper bounds\\ resources\end{tabular}                   \\ \hline
Type~3               & \multicolumn{2}{c|}{None~/~Full}                                                                                         & Required                                                                           & \begin{tabular}[c]{@{}c@{}}Lower bounds\\ resources\end{tabular}                   \\ \hline
\end{tabular}
\label{table:relation_deltar_attackers}
\end{table}
In this paper, we introduced security indices $\delta$ and $\delta_r$.
These indices can be used for localizing vulnerable actuators within the system and development of defense strategies. 
We first analyzed $\delta$, which is more suitable for small scale systems. 
A method for computing $\delta$ was derived, and it was shown that $\delta$ can potentially be increased by placing additional sensors. 
We then showed that $\delta$ may not be appropriate index for large scale networked systems since it is: (1)~NP hard to calculate; (2)~Vulnerable to system variations; (3)~Based on the assumption that the attacker knows the entire system model. 

The robust security index $\delta_r$ was then introduced as a replacement of $\delta$. 
The robust index: (1)~Can be calculated efficiently; (2)~Is robust to system variations; (3)~Can be related to both the full and limited model knowledge attackers, as summarized in Table~\ref{table:relation_deltar_attackers}.
Additionally, two sensor placement problems for increasing $\delta_r$ were proposed, and it was shown that suboptimal solutions with performance guarantees of these problems can be obtained efficiently.
Finally, the properties of $\delta$ and $\delta_r$ were illustrated through numerical examples.

The future work will go into the following directions. 
Firstly, beside perfectly undetectable attacks, there exist many other dangerous types of attacks. 
Therefore, we plan to investigate if novel types of security indices can be formulated based on these attack models.  
Secondly, the sensor placement problems considered in the paper were formulated without taking the security index $\delta$ into consideration. 
The future work will investigate if it is possible to derive sensor placement strategies for improving $\delta$ and $\delta_r$ simultaneously.

\appendix

\subsection{Proofs of Section~\ref{section:properties}} \label{appendix:proofs_1}

\hspace{-3.2mm}\textbf{Proof of Proposition~\ref{theorem:condition}.} Before we move to the proof, we introduce a sufficient and necessary condition for existence of perfectly undetectable attacks.
  \begin{lemma}\label{lemma:condition_undetectable_attacks}\cite[Theorem 1]{6816560}\cite[Theorem 7]{7479526} A perfectly undetectable attack conducted using components $\mathcal{I}_a \subseteq \mathcal{I}$ exists if and only if $\text{normrank }G^{(\mathcal{I}_a)}<|\mathcal{I}_a|$.
\end{lemma}


\textit{Proof of Proposition~\ref{theorem:condition}: }($\Rightarrow$) 
Let  $\mathcal{A}$ be the $\mathcal{Z}$--transform of $a$. Assume there exists a perfectly undetectable attack $\mathcal{A}$ with $\mathcal{A}^{(i)} \neq 0$. We split the proof into two cases. 

Case 1. Assume first $\text{normrank } G^{(\mathcal{I}_a \setminus i)} = |\mathcal{I}_a|-1$. Since undetectable attacks are possible, then it follows from Lemma~\ref{lemma:condition_undetectable_attacks}
that $\text{normrank } G^{(\mathcal{I}_a)}< |\mathcal{I}_a|$. 
On the other hand 
$$\text{normrank } G^{(\mathcal{I}_a)}\geq\text{normrank } G^{(\mathcal{I}_a \setminus i)}=|\mathcal{I}_a|-1,$$
which implies $\text{normrank } G^{(\mathcal{I}_a)} = |\mathcal{I}_a|-1$. Thus,~\eqref{eqn: condition_normal_ranks} holds.

Case 2. Assume now  $\text{normrank } G^{(\mathcal{I}_a \setminus i)} < |\mathcal{I}_a|-1$.  Let $\mathcal{I}_b \subseteq \mathcal{I}_a \setminus i$ be such that: 
\begin{itemize}
\item[(i)] The columns of $G^{(\mathcal{I}_b)}$ span the columns of $G^{(\mathcal{I}_a \setminus i)}$;
\item[(ii)] $ \text{normrank }G^{(\mathcal{I}_b)}=\text{normrank }G^{(\mathcal{I}_a \setminus i)} =|\mathcal{I}_b|$. 
\end{itemize}
Since (i) holds, we can find $\mathcal{A}'$ that satisfies $G^{(\mathcal{I}_a \setminus i)}\mathcal{A}^{(\mathcal{I}_a \setminus i)}=G^{(\mathcal{I}_b)}\mathcal{A}'.$ 
From the latter relation and $G\mathcal{A}=0$, it follows $$G\mathcal{A}=G^{(\mathcal{I}_a \setminus i)}\mathcal{A}^{(\mathcal{I}_a \setminus i)}+G^{(i)}\mathcal{A}^{(i)}=G^{(\mathcal{I}_b)}\mathcal{A}'+G^{(i)}\mathcal{A}^{(i)}=0.$$
This implies that $[(\mathcal{A}')^T \hspace{1mm }\mathcal{A}^{(i)}]^T$ is a perfectly undetectable attack against $[G^{(\mathcal{I}_b)} \hspace{1mm} G^{(i)}]$ with $\mathcal{A}^{(i)} \neq 0$. We then have
\begin{equation} \label{eqn:prop1_case21}
\begin{aligned}
\text{normrank} [G^{(\mathcal{I}_b)} \hspace{1mm} G^{(i)}]&\stackrel{*}{=}\text{normrank } G^{(\mathcal{I}_b)}\\&\stackrel{**}{=}\text{normrank }G^{(\mathcal{I}_a \setminus i)},
\end{aligned}
\end{equation}
where (*) follows from (ii) and Case 1, and (**) from (ii).
 Since $G^{(\mathcal{I}_b)}$ spans the columns of $G^{(\mathcal{I}_a \setminus i)}$, we have
\begin{equation} \label{eqn:prop1_case22}
\begin{aligned}
\text{normrank }G^{(\mathcal{I}_a)}&=\text{normrank} [G^{(\mathcal{I}_a \setminus i)}  \hspace{1mm} G^{(i)}]\\
&=\text{normrank}[G^{(\mathcal{I}_b)} \hspace{1mm} G^{(i)}].
\end{aligned}
\end{equation}
From~\eqref{eqn:prop1_case21} and~\eqref{eqn:prop1_case22}, we conclude that~\eqref{eqn: condition_normal_ranks} holds.
  
($\Leftarrow$) If~\eqref{eqn: condition_normal_ranks} holds, then there exist real rational functions $P$ and $Q\neq 0$, such that $G^{(\mathcal{I}_a\setminus i)}P + G^{(i)}Q=0. $
Thus, any attack signal $\mathcal{A}^{(i)}$ can be masked by applying $\mathcal{A}^{(\mathcal{I}_a \setminus i)}=P\mathcal{A}^{(i)}/Q$
 on the remaining attacked components.  \hfill $\blacksquare$

\hspace{-3.2mm}\textbf{Proof of Proposition~\ref{prop:adding_sensors_delta}. }
By adding a new sensor to the system, we introduce additional constraints to Problem~\ref{problem:sec_index_perfectly_undetectable}.
Thus, $\delta'(i)$$<$$\delta(i)$ cannot hold. 
If a new sensor is not protected, the attacker can gain control over it.
This can be interpreted as removing the aforementioned constraints from the problem. 
Hence, $\delta'(i)$ is at most by one larger than $\delta(i)$ in this case.  
By adding a new actuator, the number of decision variables of Problem~\ref{problem:sec_index_perfectly_undetectable} increases, and the number of constraints remains the same. 
Therefore, $\delta'(i)\leq\delta(i)$ holds.\hfill $\blacksquare$

\hspace{-3.2mm}\textbf{Proof of Theorem~\ref{theorem:NPhardness}.}
To prove NP-hardness of Problem~\ref{problem:sec_index_perfectly_undetectable}, it suffices to show that every instance of an NP--hard problem can be mapped into Problem~\ref{problem:sec_index_perfectly_undetectable}. For this purpose, we use the sparse recovery problem
\begin{equation} \label{eqn:zero_norm1}
\underset{d}{\text{minimize}} \hspace{2mm} ||d||_0 \hspace{5mm}\text{subject to }\hspace{2mm} Fd=\bar{y},
\end{equation}
where $F \in \mathbb{R}^{p \times m}$ and $\bar{y} \in  \mathbb{R}^{p}$ are given. This problem is known to be NP--hard~\cite{bruckstein2009sparse}.
Let $F$ and $\bar{y}$ be arbitrary selected. 
Set $A$$=$$\textbf{0}_{m\times m}$, $B$$=$$\textbf{I}_{m}$, $C$$=$$[-\bar{y} \hspace{1mm} F]$, $D$$=$$\textbf{0}_{p\times m}$, and $i=1$. Then $a=a_u$ and $x(k+1)=a_u(k)$.
Hence, Problem~\ref{problem:sec_index_perfectly_undetectable} becomes
\begin{equation} \label{eqn:zero_norm2}
\underset{a_u}{\text{minimize}}\hspace{1mm}||a_u||_{0}\hspace{3mm}\text{subject to}  \hspace{2mm}Ca_u(k)=0, \hspace{1mm}   a_u^{(1)}\neq 0.
\end{equation}

It can be seen that to solve~\eqref{eqn:zero_norm2} for all $k$, it suffices to solve it for a single $k$. 
 Thus,~\eqref{eqn:zero_norm2} reduces to  
\begin{equation*}
\underset{a_u(0)}{\text{minimize}} \hspace{1mm} ||a_u(0)||_{0}\hspace{3mm}
\text{subject to}  \hspace{2mm} Ca_u(0)=0,\hspace{1mm}a_u^{(1)}(0)= 1,
\end{equation*}
where the substitution of $a_u^{(1)}(0)$$ \neq$$ 0$ with $a_u^{(1)}(0)$$=$$1$ is without loss of generality. 
Let $a_u(0)$$=$$[1 \hspace{1mm}d^T]^T$. 
Then minimizing $||a_u(0)||_{0}$ is equivalent to minimizing $||d||_{0}$, which is the objective function of~\eqref{eqn:zero_norm1}. 
Moreover, we also have that $Ca_u(0)=[-\bar{y} \hspace{1mm} F]a_u(0)$$=$$-\bar{y} +Fd.$ 
Thus, $Ca_u(0)=0$ implies $Fd=\bar{y},$ which is the constraint of~\eqref{eqn:zero_norm1}. 
Therefore, every instance of the NP--hard problem~\eqref{eqn:zero_norm1} can be mapped into Problem~\ref{problem:sec_index_perfectly_undetectable}, which concludes the proof. \hfill $\blacksquare$

\subsection{Proofs of Section~\ref{section:upper_bound}} \label{appendix:proofs_2}

\hspace{-3.2mm}\textbf{Proof of Theorem~\ref{proposition:sec_index_sufficient_condition}.}
Let $\mathcal{X}_a \cup \mathcal{Y}_a$ be a vertex separator of $u_i$ and $t$ in the graph $\mathcal{G}_t$. To prove the claim, we introduce an attack strategy that uses the components $\mathcal{U}_a$ and $\mathcal{Y}_a$. We then prove that this strategy is actively using $u_i$, and it is perfectly undetectable in any realization $A,B,C$. 

For actuator $u_i$, the attacker injects an arbitrary signal $a^{(u_i)} \neq 0$.
This ensures that $u_i$ is used in the attack actively. 
For other actuators $u_j \in \mathcal{U}_a \setminus u_i$, the attack is 
\begin{equation}\label{ref:attackauj}
a^{(u_j)}(k)=-A(p,:)x(k)/B(p,j),
\end{equation} 
where $A{(p,:)}$ is the row of $A$ corresponding to attacked actuator $u_j$, and $B(p,j)$ is the non-zero element of $B$ multiplying $u_j$ (such element exists for any realization due to Assumption~\ref{assumption:BandCmatrix}.(3)). 
For $y_l \in \mathcal{Y}_a$, the attack is 
\begin{equation}\label{ref:attackayj}
a^{(y_l)}(k)=-C(l,:)x(k), 
\end{equation}
where $C(l,:)$ represents the row of $C$ corresponding to $y_l$. 
For the attacker with the full model knowledge, this strategy can be constructed for any realization.
Namely, he/she knows the values for $A(p,:)$,$B(p,j)$,$C(l,:)$, and can predict the value of $x(k)$ for any $k\in\mathbb{Z}_{\geq 0}$ based on the model and the attack signals.
We now prove that this strategy is perfectly undetectable, that is, $y=0$.

We first consider attacked sensors. 
For any $y_l \in \mathcal{Y}_a$ and $k\in\mathbb{Z}_{\geq 0}$, we have $y_l(k)=C(l,:)x(k)+a^{(y_l)}(k)\stackrel{\eqref{ref:attackayj}}{=}0$. 
Thus, the attacked measurements are equal to 0. 
It remains to be shown that the non-attacked measurements are also 0. 

Consider first the non-attacked sensors measuring the states from $\mathcal{X}_a$.
Let $x_p \in \mathcal{X}_a$, and let $u_j\in \mathcal{U}_a \setminus u_i$ be adjacent to $x_p$. 
Then $x_p(k+1)=A(p,:)x(k)+B(p,j)a^{(u_j)}(k)\stackrel{\eqref{ref:attackauj}}{=}0$.  
Thus, the non-attacked measurements of the states from $\mathcal{X}_a$ are 0. 
Let now $\mathcal{X}_b$ be the set of all the states for which there exists a directed path from $u_i$ that does not contain the states from $\mathcal{X}_a$. 
These states cannot be measured using the non attacked sensors.
That would imply that there exists a directed path in between $u_i$ and $t$ not intersected by $\mathcal{X}_a\cup \mathcal{Y}_a$, which is  in contradiction with the assumption that $\mathcal{X}_a\cup \mathcal{Y}_a$ is a vertex separator of $u_i$ and $t$. 
Finally, let $\mathcal{X}_c=\mathcal{X} \setminus (\mathcal{X}_b \cup \mathcal{X}_a)$ be the set of all the remaining states.  
Note that the directed edges $(x_b,x_c)$, $x_b \in \mathcal{X}_b$, $x_c \in \mathcal{X}_c$, cannot exist.
That would imply that there exists a directed path from $u_i$ to $x_c$ that does not contain the states from $\mathcal{X}_a$, so $x_c$ would belong to $\mathcal{X}_b$. 
Thus, the states from $\mathcal{X}_c$ cannot be directly influenced by the states from $\mathcal{X}_b$.  
Since $x(0)=0$, $u=0$, and the states $\mathcal{X}_a$ are equal to 0, we conclude that the states $ \mathcal{X}_c$ also remain equal to 0 during the attack. 
Thus, the non-attacked measurements of these states remain 0. 
With this, we prove that all of the non-attacked measurements are equal to 0, so the attack strategy is perfectly undetectable. \hfill $\blacksquare$

\subsection{Proofs of Section~\ref{section:upper_bound_properties}} \label{appendix:proofs_3}

\hspace{-3.2mm}\textbf{Proof of Proposition~\ref{theorem:security_index_ts_cut}. }Statement~1.  
Let $\mathcal{U}_a \cup \mathcal{Y}_a$ be a solution of Problem~\ref{problem:sec_index_upper_bound}, and $\mathcal{X}_a \cup \mathcal{Y}_a$ be a corresponding vertex separator. 
Let $\mathcal{E}_c \subseteq \mathcal{E}_{u_i}$ be constructed as follows. 
For each $x_k \in \mathcal{X}_a$, we add $(x_{k_{in}},x_{k_{out}})$ to $\mathcal{E}_c$. For each $y_j \in \mathcal{Y}_a$ with $(x_{k},y_j) \in \mathcal{E}_{xy}$, we add $(x_{k_{out}},t)$ (resp. $(x_{k},t)$) to $\mathcal{E}_c$ if $x_k$ is Type~1 (resp. Type~2). 
If there exists more than one measurement of $x_{k}$, then all of them must belong to $\mathcal{Y}_a$.
Otherwise, there would exist a path from $u_i$ to $t$ not intersected by  $\mathcal{X}_a \cup \mathcal{Y}_a$, or $y_j$ would not be a part of an optimal solution.  
  It follows from the construction of $\mathcal{G}_{u_i}$ that the edges added to $\mathcal{E}_c$ have the cost $\delta_c=$$|\mathcal{U}_a\setminus i|$$+|\mathcal{Y}_a|$$=\delta_r(u_i)-1.$
We now show that $\mathcal{E}_c$ is an edge separator of $u_i$ and $t$ in $\mathcal{G}_{u_i}$ (Claim~1) of the minimum cost (Claim~2).
This implies $\delta_r(u_i)=$$\delta_c+1$$=\delta^*+1$, and proves Statement~1. 

Claim~1. 
Assume $\mathcal{E}_c$ is not an edge separator. 
Then there exists a simple directed path $u_i,x_{j_1},\ldots,x_{j_n},t$ (Path~1) in $\mathcal{G}_{u_i}$, which is not intersected by $\mathcal{E}_c$.
By the construction of $\mathcal{G}_{u_i}$, that implies that there exists a simple directed path $u_i,x_{k_1},$$\ldots$$,x_{k_m},y_l,t$ (Path~2) in $\mathcal{G}_t$, obtained from Path~1 by replacing every pair $x_{p_{in}},x_{p_{out}}$ that corresponds to $x_p$ of Type~1 by $x_p$, and by inserting a measurement $y_l$ of $x_{k_m}$. 
Path~2 has to be intersected with $\mathcal{X}_a \cup \mathcal{Y}_a$. 
Then either exists $x_p\in \mathcal{X}_a$ that belongs to Path~2 or $y_l \in $$\mathcal{Y}_a$.
However, then either ($x_{p_{in}},$$x_{p_{out}}$) or $(x_{j_n}$$,t)$ belongs to $\mathcal{E}_c$. 
This contradicts existence of Path~1, so Claim~1 holds.  

Claim~2. Assume there exist an edge separator $\mathcal{E}'_c$ with the cost $\delta'<\delta_c$. 
Let $\mathcal{U}'_a \cup \mathcal{Y}'_a$ be constructed as follows.
 For each $(x_{k_{in}},x_{k_{out}})$ from $\mathcal{E}_c$, we add $u_j$ to $\mathcal{U}_a'$, where $u_j$ is adjacent to $x_k$.
For each edge $(x_{p_{out}},t)$ or $(x_{p},t)$ from $\mathcal{E}_c$, we add all the measurements of $x_p$ to $\mathcal{Y}_a'$. 
All of these measurements must be unprotected (otherwise $\delta'=+\infty>\delta_c$).
We add $u_i$ to $\mathcal{U}'_a$. 
Note that $\mathcal{E}'_c$ cannot contain edges of other types, because their weight is $+\infty$, which would imply $\delta'>\delta_c$.

Firstly, we prove that $\mathcal{U}'_a \cup \mathcal{Y}'_a$ must be a feasible point of Problem~\ref{problem:sec_index_upper_bound}.
Assume that is not the case. 
Since, $u_i \in \mathcal{U}'_a$ and all the measurements from $\mathcal{Y}'_a$ are unprotected, it follows
that there exists a simple directed path $u_i,x_{k_1},$$\ldots$$,x_{k_m},y_l,t$ (Path~1') in $\mathcal{G}_{t}$, in which none of the states are adjacent to $\mathcal{U}'_a \setminus u_i$, and $y_l \notin \mathcal{Y}'_a$. 
That implies that there exists a simple directed path in $\mathcal{G}_{u_i}$ obtained from Path~1' by replacing each node $x_p$ of Type~1 from this path by $x_{p_{in}},x_{p_{out}}$, and removing $y_l$. 
By the construction of $\mathcal{U}'_a \cup \mathcal{Y}'_a$ and $\mathcal{G}_{u_i}$, this path cannot be intersected by $\mathcal{E}'_c$. 
This would contradict the assumption that $\mathcal{E}'_c$ is an edge separator, so $\mathcal{U}'_a \cup \mathcal{Y}'_a$ has to be a feasible point of Problem~\ref{problem:sec_index_upper_bound}.
However, then $\mathcal{U}_a \cup \mathcal{Y}_a$ is not a solution of Problem~\ref{problem:sec_index_upper_bound} because
$|\mathcal{U}'_a \cup \mathcal{Y}'_a|=\delta'+1<|\mathcal{U}_a \cup \mathcal{Y}_a|=\delta_c+1$.
Thus, $\mathcal{E}'_c$ cannot exist, and Claim~2 holds. 

Statement~2. In this case, there has to exist a simple directed path $u_i,x_{j_1},$$\ldots$$,x_{j_n},y_l,t$ in $\mathcal{G}_{t}$ that contains only Type~2 states and protected measurement $y_l$. 
Then the path $u_i,x_{j_1},\ldots,x_{j_n},t$ exists in $\mathcal{G}_{u_i}$, and the weights of all the edges from this path are  $+\infty$.  
Any edge separator needs to cut this path, which implies $\delta^*=+\infty$. 
 \hfill $\blacksquare$
 
\hspace{-3.2mm}\textbf{Proof of Proposition~\ref{prop:FMNA_resources}. }
Let $\mathcal{X}_a$ be defined as in~\eqref{eqn:set_xa}. 
We prove the claim by showing that
$\mathcal{X}_a\cup \mathcal{Y}_a$ has to be a vertex separator of $u_i$ and $t$ in $\mathcal{G}_t$. 
Assume this is not the case.
Then there exists at least one simple directed path  $u_i,x_{i_0},\ldots,x_{i_n},y_{l},t$ (Path~1) not intersected by $\mathcal{X}_a \cup \mathcal{Y}_a$.
We now show that this implies existence of at least one realization of the structural model $[A],[B],[C]$ in which a perfectly undetectable attacks against $u_i$ cannot be conducted. 

Assume the following realization of matrices $A$ and $C$. 
For $x_{i_0}$ from Path~1, $A(i_0,:)=0$. 
This ensures that $x_{i_0}$ cannot be influenced by any state.
For any other $x_{i_k}$ from Path~1, $A(i_k,j)\neq0$ (resp. $A(i_k,j)=0$) if $j= i_{k-1}$ (resp. $j\neq i_{k-1}$).  
This guarantees that the only state that influences $x_{i_{k}}$ is $x_{i_{k-1}}$. 
For edge $(x_{i_n},y_{l})\in \mathcal{E}_{xy}$ from Path~1, $C(l,{i_n})\neq 0$.
This ensures that $y_{l}(k)\neq 0$ once $x_{i_n}(k)\neq 0$.
We now show that if this realization is present, a perfectly undetectable attack in which $u_i$ is actively used does not exist. 

Let $a^{(u_i)} \neq 0$ be an arbitrary attack signal against $u_i$, and let $k_0$ be the first time instant  for which 
$a^{(u_i)}(k_0)\neq 0$. 
Since  $u=0$ and $a^{(u_i)}$ is the only attack signal that can directly influence $x_{i_0}$ (due to Assumptions~\ref{assumption:BandCmatrix}.(1) and~\ref{assumption:BandCmatrix}.(2)), we have
$x_{i_0}(k_0+1)=A(i_0,:)x(k_0)+B(i_0,i)a^{(u_i)}(k_0).$ 
Given that  $A(i_0,:)=0$ and $B(i_0,i) \neq 0$ (Assumption~\ref{assumption:BandCmatrix}.(3)), it follows $x_{i_0}(k_0+1)\neq0$. 
We now show $x_{i_1}(k_0+2)\neq 0$.
Note that the only state that influences $x_{i_1}$ is $x_{i_0}$. 
Moreover, since $x_{i_1}$ cannot be influenced by attacked actuators ($x_{i_1}\notin \mathcal{X}_a$), and $u=$$0$,  it follows $x_{i_1}(k_0+2)=A(i_1,i_0)x_{i_0}(k_0+1)\neq0.$ By applying the similar reasoning to all other states from Path~1, it can be shown that $x_{i_n}(k_0+n+1) \neq 0$.
Thus, $y_{l}(k_0+n+1)\neq 0$, which implies that the attack is revealed. 
Since $a^{(u_i)}$ was arbitrary selected,  there exists no perfectly undetectable attacks with $u_i$ actively used in this realization.  

This contradicts the assumption that the attacker can conduct a perfectly undetectable attack against $u_i$ in any realization of $[A],[B],[C]$ by using $\mathcal{U}_a$ and $\mathcal{Y}_a$. 
Thus, $\mathcal{X}_a\cup \mathcal{Y}_a$ has to be a vertex separator of $u_i$ and $t$ in $\mathcal{G}_t$. 
Since $\delta_r(u_i)$ is the minimum number of attacked sensors and actuators that ensures $\mathcal{X}_a\cup \mathcal{Y}_a$ is a vertex separator of $u_i$ and $t$ with $u_i \in \mathcal{U}_a$, the claim of the proposition holds. 
\hfill $\blacksquare$

\hspace{-3.2mm}\textbf{Proof of Proposition~\ref{theorem:restricted_attacker}.}
($\Rightarrow$) The proof is by contradiction. 
If $\mathcal{X}_a \cup \mathcal{Y}_a$ is not a vertex separator of $u_i$ and $t$ in $\mathcal{G}_t$,  we know from the proof of Proposition~\ref{prop:FMNA_resources} that we can find at least one realization in which it is not possible to conduct a perfectly undetectable attack against $u_i$. 
Thus, $\mathcal{X}_a \cup \mathcal{Y}_a$ has to be a vertex separator of $u_i$ and $t$. 

($\Leftarrow$) If $\mathcal{X}_a \cup \mathcal{Y}_a$ is a vertex separator of $u_i$ and $t$, the attacker can conduct a perfectly undetectable attack against $u_i$ using the strategy similar to the one in the proof of Theorem~\ref{proposition:sec_index_sufficient_condition}. 
For actuator $u_i$, the attacker injects an arbitrary signal $a^{(u_i)} \neq 0$. 
For other actuators $u_j $$\in$$ \mathcal{U}_a $$\setminus $$u_i$ with $(u_j,x_p) $$\in$$ \mathcal{E}_{ux}$, the attack is given by 
$a^{(u_j)}(k)$$=$$-A(p,:)x(k)/B(p,j)$.
For $y_l \in \mathcal{Y}_a$, the attacker selects $a^{(y_l)}(k)$ to maintain $y_l(k)=0$.

The Type~2 attacker can construct this attack.
Firstly, the attacker knows the values for $A(p,:)$,$B(p,:)$ that correspond to actuators $u_j$$ \in$$ \mathcal{U}$$\setminus$$ u_i$.
Secondly, the attacker can construct $A(p,:)x(k)$, since he/she knows the values of in-neighbors of $x_p$, while the elements of $A(p,:)$ that correspond to other states are equal to 0.
Thirdly, the Type~2 attacker can also set the signals of attacked sensors and actuators to an arbitrary value, so he/she can maintain $y_l(k)$$=$$0$. 
The proof that $y$$=$$0$ is then analogous to the proof of Theorem~\ref{proposition:sec_index_sufficient_condition}.  
\hfill $\blacksquare$

\hspace{-3.2mm}\textbf{Proof of Proposition~\ref{prop:WMNA_resources}.}
We prove the claims by showing that $\mathcal{Y}_a$ has to be a vertex separator of $u_i$ and $t$ in $\mathcal{G}_t$.
Namely, existence of a path from $u_i$ to $t$ in $\mathcal{G}_t$ implies that there exist at least one sensor $y_j$ that is not compromised by the attacker.
From the proof of Proposition~\ref{prop:FMNA_resources}, we know that there exists at least one realization of the system in which the attack against $u_i$ triggers $y_j$.
Since the Type~3 attacker has knowledge of only $[A],[B],[C]$, he/she does not know if the attack  against $u_i$ would be visible in $y_j$ or not.
Thus, the Type~3 attacker needs to attack $y_j$ to ensure being perfectly undetectable.
Therefore, $\mathcal{Y}_a$ has to form a vertex separator of $u_i$ and $t$. 
By the definition, $\delta_r(u_i)-1$ is the size of the minimum vertex separator of $u_i$ and $t$ in $\mathcal{G}_t$ (we subtract 1 from  $\delta_r(u_i)$ to exclude $u_i$).
Hence, $|\mathcal{Y}_a|\geq \delta_r(u_i)-1$. 
Finally, if $\delta_r(u_i)=+\infty$, then there exists a path in between $u_i$ and a protected sensor.
This implies that $\mathcal{Y}_a$ cannot be a vertex separator.
Hence, the Type~3 attacker cannot ensure that a perfectly undetectable attack against $u_i$ remains perfectly undetectable, because he/she does not know if the aforementioned protected sensor would be triggered.  
\hfill $\blacksquare$

\subsection{Proofs of Section~\ref{section:improving_upper_bound}} \label{appendix:proofs_4}

\hspace{-3.2mm}\textbf{Proof  of Theorem~\ref{proposition:place_to_put_sensors}.}
Assume we place $y_{l}$ to monitor any of the states from $\mathcal{X}_{u_i}$.
We then introduce at least one additional directed path $u_i,x_j,\ldots,y_l,t$ from $u_i$ to $t$, which does not contain states adjacent to $\mathcal{U}\setminus u_i$. 
Thus, the only way to remove this path is by adding $y_{l}$ to a new vertex separator.
If $y_{l}$ is protected, that is not possible, so $\delta_r'(u_i)=+\infty$. 
Otherwise, the attacker must attack $y_{l}$, thus  $\delta_r'(u_i)=\delta_r(u_i)+1$. 

We now show that if for every $x_j$$ \in$$ \mathcal{X}$ for which there exists $(u_k,x_j) $$\in $$ \mathcal{E}_{ux}$, there also exists $(x_j,y_p)$$\in$$  \mathcal{E}_{xy}$, then the only way to improve $\delta_r(u_i)$ is by placing sensors within $\mathcal{X}_{u_i}$. 
Let $\mathcal{U}_a $$\cup $$\mathcal{Y}_a$ be a solution of Problem~\ref{problem:sec_index_upper_bound} for $u_i$. 
We first form another optimal solution $\mathcal{U}_a' $$\cup $$\mathcal{Y}_a'$ from $\mathcal{U}_a$$ \cup$$ \mathcal{Y}_a$. 
The set $\mathcal{Y}_a'$ is formed by removing from $\mathcal{Y}_a$ any $y_j$ which measures $x_k $$\in$$ \mathcal{X}$ that is adjacent to $u_l$$ \in$$ \mathcal{U} $$\setminus $$u_i$. 
As a substitute of $y_j$, we add $u_l$ to $\mathcal{U}_a'$. 
We then add all the actuators $\mathcal{U}_a$ to $\mathcal{U}_a'$. 
This ensures that for all the states that are both directly influenced by an actuator and measured by a sensor, we always select an actuator to belong to a solution of Problem~\ref{problem:sec_index_upper_bound} rather than a sensor.  
Finally, let $\mathcal{X}_a' $ be defined as in~\eqref{eqn:set_xa} based on $\mathcal{U}_a' $.

Let a sensor be placed to measure $x_{l}$$ \notin$$ \mathcal{X}_{u_i}$.
If there are no directed paths from $u_{i}$ to $x_{l}$, or if all the paths from $u_{i}$ to $x_{l}$ are intersected by $\mathcal{X}_a' $$\cup$$ \mathcal{Y}_a'$, then $\mathcal{U}_a' $$\cup $$\mathcal{Y}_a'$ is still a solution of Problem~\ref{problem:sec_index_upper_bound} and $\delta_r(u_i)$ is not increased.
Thus, assume there exist a simple directed path $u_{i},\ldots,x_{l}$ (Path~1) not intersected by $\mathcal{X}_a' $$\cup$$ \mathcal{Y}_a'$. 
Since $x_{l}$$ \notin $$\mathcal{X}_{u_i}$, there has to exist at least one state $x_p$ from Path~1 adjacent to an actuator.  
Then $x_p$ has to be also adjacent to a sensor, which implies existence of a directed path in between $u_i$ and $t$ passing through $x_p$ that is not intersected by $\mathcal{X}_a' \cup \mathcal{Y}_a'$.
This is not possible, since $\mathcal{U}_a' \cup \mathcal{Y}_a'$ is a solution of Problem~\ref{problem:sec_index_upper_bound}. 
Hence, Path~1 cannot exists.
Therefore, we cannot increase $\delta_r(u_i)$ by placing sensors outside $\mathcal{X}_{u_i}$. 
\hfill $\blacksquare$


\hspace{-3.2mm}\textbf{Proof of Proposition~\ref{theorem:improving_bound_subm_2}.}
We first show that $g_{u_i}$ is submodular, nondecreasing, and integer-valued. 
Firstly, 
$
w_{y_j}=|x_{y_j} \cap \mathcal{X}_{u_i}|
$ 
is a binary integer constant. 
Thus, $g_l(\mathcal{Y}_p)=\sum_{y_j \in \mathcal{Y}_p}w_{y_j}$ is a linear function, so it is   both submodular~\cite[Section 2]{bach2013learning} and nondecreasing (sum of nonnegative numbers).
Since we have $g_{u_i}(\mathcal{Y}_p)$$=\min\{ g_l(\mathcal{Y}_p),k_{u_i}\}$, it follows from Lemma~\ref{lemma:properties_submodular} that $g_{u_i}$ is submodular and non-decreasing. 
Function $g_{u_i}$ is also integer valued, since $g_l$ and $k_{u_i}$ are integer valued.
Thus, it follows from Lemma~\ref{lemma:properties_submodular} that $G$ is submodular, nondecreasing, and integer valued. 
We also have $G(\emptyset)=0$, which implies that $G$ has the same properties as the set function from~\eqref{eqn:sub_problem_2}.
Thus, the claim of the proposition hold.  \hfill $\blacksquare$

\hspace{-3.2mm}\textbf{Proof of Proposition~\ref{theorem:improving_bound_subm_1}.}
The function $g'_{u_i}$ is known to be submodular~\cite[Section 2]{bach2013learning}. 
Additionally, $g'_{{u_i}}$ is a nondecreasing function, since $|\mathcal{X}_p \cap \mathcal{X}_{u_i}|$ is nondecreasing  in $\mathcal{X}_p$. 
We then have from Lemma~\ref{lemma:properties_submodular} that $G'$ is submodular and nondecreasing.
In addition, $G'(\emptyset)=0$.
Hence, $G'$ has the same properties as the function from~\eqref{eqn:sub_problem_1}, which concludes the proof. 
\hfill $\blacksquare$

\bibliographystyle{IEEEtran}
\bibliography{autosam} 

\newpage
\vskip -2\baselineskip plus -1fil
\begin{IEEEbiography}
{Jezdimir Milo\v{s}evi\'{c}} received his M.Sc. degree in Electrical Engineering
and Computer Science in 2015 from the School of Electrical Engineering, University of Belgrade, Serbia.
He is currently pursuing the Ph.D. degree at the
Department of Automatic Control, KTH Royal Institute of Technology, Sweden. 
He was a visiting researcher at the University of Hawaii at Manoa in 2014, and Massachusetts Institute of Technology in 2018.
His research interests are within cyber-security of industrial control systems.
\end{IEEEbiography}
\vskip -2\baselineskip plus -1fil
\begin{IEEEbiography}
{Andr\'{e} Teixeira } is an Associate Senior Lecturer at the Division of Signals and Systems, Department of Engineering Sciences, Uppsala University, Sweden. He received the M.Sc. degree in electrical and computer engineering from the Faculdade de Engenharia da Universidade do Porto, Porto, Portugal, in 2009, and the Ph.D. degree in automatic control from the KTH Royal Institute of Technology, Stockholm, Sweden, in 2014. From 2014 to 2015, he was a Postdoctoral Researcher at the Department of Automatic Control, KTH Royal Institute of Technology, Stockholm, Sweden. From October 2015 to August 2017, he was an Assistant Professor at the Faculty of Technology, Policy and Management, Delft University of Technology.
\end{IEEEbiography}
\vskip -2\baselineskip plus -1fil
\begin{IEEEbiography}
{Karl Henrik Johansson }
is Director of the Stockholm Strategic Research Area ICT The Next Generation and Professor at the School of Electrical Engineering and Computer Science, KTH Royal Institute of Technology. He received MSc and PhD degrees from Lund University. He has held visiting positions at UC Berkeley, Caltech, NTU, HKUST Institute of Advanced Studies, and NTNU. His research interests are in networked control systems, cyber-physical systems, and applications in transportation, energy, and automation. He is a member of the IEEE Control Systems Society Board of Governors, the IFAC Executive Board, and the European Control Association Council. He has received several best paper awards and other distinctions. He has been awarded Distinguished Professor with the Swedish Research Council and Wallenberg Scholar. He has received the Future Research Leader Award from the Swedish Foundation for Strategic Research and the triennial Young Author Prize from IFAC. He is Fellow of the IEEE and the Royal Swedish Academy of Engineering Sciences, and he is IEEE Distinguished Lecturer. 
\end{IEEEbiography}
\vskip -2\baselineskip plus -1fil
\begin{IEEEbiography}
{Henrik Sandberg } is Professor at the Department of Automatic Control, KTH Royal Institute of Technology, Stockholm, Sweden. He received the M.Sc. degree in engineering physics and the Ph.D. degree in automatic control from Lund University, Lund, Sweden, in 1999 and 2004, respectively. From 2005 to 2007, he was a Post-Doctoral Scholar at the California Institute of Technology, Pasadena, USA. In 2013, he was a visiting scholar at the Laboratory for Information and Decision Systems (LIDS) at MIT, Cambridge, USA. He has also held visiting appointments at the Australian National University and the University of Melbourne, Australia. His current research interests include security of cyber-physical systems, power systems, model reduction, and fundamental limitations in control. Dr. Sandberg was a recipient of the Best Student Paper Award from the IEEE Conference on Decision and Control in 2004, an Ingvar Carlsson Award from the Swedish Foundation for Strategic Research in 2007, and Consolidator Grant from the Swedish Research Council in 2016. He has served on the editorial board of IEEE Transactions on Automatic Control and is currently Associate Editor of the IFAC Journal Automatica.
\end{IEEEbiography}

\end{document}